\documentclass[12pt]{article} 
\pdfoutput=1
\usepackage{jheppub}
\usepackage{amsfonts}
\usepackage{amscd}
\usepackage{amssymb}
\usepackage{amsmath}
\usepackage{epsfig}
\usepackage{latexsym}

\def \be  {\begin{equation}}
\def \ee  {\end{equation}}
\def \ba  {\begin{eqnarray}}
\def \ea  {\end{eqnarray}}
\def \bb  {\begin{thebibliography}}
\def \eb  {\end{thebibliography}}
\def \lab #1 {\label{#1}}

\newcommand\PT{\mathbb{PT}}
\newcommand\cA{\mathcal{A}}

\newcommand\cD{\mathcal{D}}

\newcommand\cL{\mathcal{L}}
\newcommand\cM{\mathcal{M}}
\newcommand\cO{\mathcal{O}}

\newcommand\cV{\mathcal{V}}
\newcommand\cW{\mathcal{W}}

\newcommand\cZ{\mathcal{Z}}
\newcommand{\p}{\partial}
\newcommand\A{\mathbb{A}}
\newcommand\PA{P\mathbb{A}}

\newcommand\T{\mathbb{T}}
\newcommand\cN{\mathcal{N}}
\newcommand\C {\mathbb{C }}

\newcommand\R {\mathbb{R }}
\newcommand\bP{\mathbb{P}}
\newcommand\CP {\mathbb{CP}}

\newcommand\Z{\mathbb{Z}}

\newcommand\rd{\mathrm{d}}

\newcommand\e{\mathrm{e}}
\newcommand\im{\mathrm{i}}

\newcommand\tr{\mathrm{tr}}

\newcommand\la{\langle}
\newcommand\ra{\rangle}
\newcommand\del{\partial}
\newcommand\delbar{\bar{\partial}}

\newcommand\hook { \lrcorner \,}

\newtheorem{thm}{Theorem}

\title{Ambitwistor strings and the scattering equations}

\author[*]{Lionel Mason}
\affiliation[*]{The Mathematical Institute,\\
Andrew Wiles Building,\\
Woodstock Road, Oxford OX2 6GG,\\ 
United Kingdom\vspace{0.2cm}}

\author[\dagger]{and David Skinner}

\affiliation[\dagger]{Department of Applied Mathematics and Theoretical Physics,\\
Wilberforce Road, Cambridge CB3 0WA,\\
United Kingdom}

\abstract{
We show that  string theories admit chiral infinite tension analogues in which only the massless parts of the spectrum survive.  Geometrically they describe holomorphic maps to spaces of complex null geodesics, known as ambitwistor spaces.   They have the standard critical space--time dimensions of string theory (26 in the bosonic case and 10 for the superstring).  Quantization leads to the formulae for tree--level scattering amplitudes of massless particles found recently by Cachazo,  He and Yuan. These representations localize the vertex operators to solutions of the same equations found by Gross and Mende to govern the behaviour of strings in the limit of high energy, fixed angle scattering. Here, localization to the scattering equations emerges naturally as a consequence of working on ambitwistor space. The worldsheet theory suggests a way to extend these amplitudes to spinor fields and to loop level.  We argue that this family of string theories is a natural extension of the existing twistor string theories.
}

\notoc

\begin{document}
\maketitle



\section{Introduction}
\label{sec:intro}

Witten's twistor string theory \cite{Witten:2003nn}  led to a
strikingly compact formula~\cite{Roiban:2004yf} for tree--level
scattering amplitudes in four-dimensional Yang-Mills theory in terms
of an integral over the moduli space of holomorphic curves in twistor
space.  More recently, analogous expressions have been found for
$\cN=8$ supergravity~\cite{Cachazo:2012da,Cachazo:2012kg,
  Cachazo:2012pz} and for ABJM theory~\cite{Huang:2012vt}.  This year,
in the remarkable series of papers
\cite{Cachazo:2013iaa,Cachazo:2013gna,Cachazo:2013hca,Cachazo:2013iea},
Cachazo, He and Yuan have presented analogous formulae based on the
ideas in \cite{Cachazo:2012da}, but now extended to describe
scattering of massless particles of spins 0, 1 or 2 in arbitrary
dimension. A striking property of these new expressions is that they
provide one of the most concrete expressions to date of the Kawai,
Lewellen and Tye notion of gravitational amplitudes being the square
of Yang-Mills amplitudes~\cite{Kawai:1985xq}, and are also closely
related to the duality between colour and kinematics found by Bern,
Carrasco and Johannson~\cite{Bern:2010ue}.

The formulae of Cachazo {\it et al.} are based on holomorphic maps of a Riemann sphere into complex momentum space
\be
P(\sigma)=\sum_{j=1}^n\frac{k_j}{\sigma-\sigma_j} :\CP^1 \rightarrow  \C^d\, , 
\ee
where the $k_j$ are the null momenta of the $n$ particles taking part in the scattering process, and the $\sigma_j$ are $n$ points on the Riemann sphere.  These points are not arbitrary, but are determined in terms of the external kinematics by imposing the scattering equations
\be\label{scatt}
k_i\cdot P(\sigma_i)=\sum_{j\neq i} \frac{k_i\cdot k_j}{\sigma_i-\sigma_j}=0\ .
\ee 
These equations were first obtained by Gross and Mende~\cite{Gross:1987ar,Gross:1987kza}, where they were shown to govern the string path integral in the limit of high energy scattering at fixed angle ($s\gg1/\alpha'$). They also underpin the twistor string formulae of \cite{Roiban:2004yf}, as first observed by Witten in \cite{Witten:2004cp}.  This is quite remarkable, since the twistor string contains only massless states and is weakly coupled suggestive of a $\alpha'\rightarrow 0$ limit rather than  $\alpha'\rightarrow\infty$.

Witten's original twistor string (with an alternative formulation by Berkovits \cite{Berkovits:2004hg} and a heterotic formulation in \cite{Mason:2007zv})  was discovered to be equivalent to a certain unphysical non-minimal version of conformal supergravity~\cite{Berkovits:2004jj} coupled to $\cN=4$ Yang-Mills. More recently, the gravitational amplitudes found in~\cite{Cachazo:2012kg} were discovered to arise from a new twistor string theory \cite{Skinner:2013xp} for $\cN=8$ supergravity.  These twistor strings are specific to these theories and it remains unclear how to extend them to other theories, or whether either has any validity for loop amplitudes.  In general one would like to be able to construct analogous string theories for more generic field theories and to have some reasonable expectation that they will, at least in favourable circumstances,  lead to the correct loop amplitudes. 

\medskip

In this paper we present a new family of string theories that are better placed to fulfill these aims and that underpin the more recent formulae of Cachazo {\it et al.}.  To motivate these theories, consider the standard first--order worldline action for a massless particle traversing a $d$ dimensional space--time $(M,g)$\footnote{These expressions are given for flat space. For a general metric $g$ the transformations involve the Christoffel symbols as generated by~\eqref{spray}.}
\be\label{null-particle}
S[X,P]=\frac{1}{2\pi}\int P_\mu\rd X^\mu -  \frac{e}{2} P_\mu P^\mu\, .
\ee
In this action, the einbein $e$ is a Lagrange multiplier enforcing the constraint $P^2=0$, and is also the worldline gauge field for the gauge transformations
\be
	\delta X^\mu = \alpha P^\mu\qquad \delta P_\mu=0 \qquad \delta e=\rd\alpha
\ee
conjugate to this constraint. We learn that $P$ must be null and that we should consider fields $X$ and $X'$ that differ by translation along a null direction to be equivalent. Consequently, the solutions to the field equations modulo this gauge redundancy are null geodesics in space--time, parametrized by the scaling of $P$. The quantization of this action leads to the massless Klein-Gordon equation.

The new chiral string theories we study may be viewed as a natural analogue of~\eqref{null-particle}, obtained by complexifying the worldline to a Riemann surface $\Sigma$ and likewise complexifying the target space so that the $X^\mu$ are holomorphic coordinates on a complexified space--time with holomorphic metric $g$.  In the simplest case, we merely replace $\rd X$ in~\eqref{null-particle} by $\delbar X = \rd\bar\sigma\, \del_{\bar\sigma} X$ to obtain the bosonic action
\be\label{boson-str}
S[X,P]=\frac1{2\pi}\int_\Sigma  P_\mu \delbar X^\mu - \frac{e}{2} P_\mu P^\mu\, .
\ee
For the kinetic term of~\eqref{boson-str} to be meaningful, we must interpret $P_\mu$ not as a scalar field, but as a complex (1,0)-form on the worldsheet, so that (suppressing the target space index) $P = P_{\sigma}(\sigma)\rd\sigma$ in terms of some local holomorphic worldsheet coordinate $\sigma$. It then follows that $e$ must now be a (0,1)-form on $\Sigma$ with values in $T\Sigma$ -- in other words a Beltrami differential.

It is perhaps not surprising that we find in section~\ref{sec:boson} that the spectrum of the string theory based on~\eqref{boson-str} contains only massless particles. 
Indeed, as we show in appendix~\ref{sec:alpha0}, \eqref{boson-str} may also be obtained by taking the $\alpha'\to0$ limit of the conventional bosonic string in a chiral way, so the usual string excitations decouple (the tachyon is also absent). However, the geometrical interpretation is quite different from that of the ordinary string.  The constraint $P^2=0$ (as a quadratic differential) and 
corresponding gauge freedom 
\be
	\delta X^\mu = \alpha \,P^\mu\qquad \delta P_\mu=0 \qquad \delta e = \delbar \alpha
\ee
survive in this model, again provided we interpret $\alpha$ as transforming as a worldsheet holomorphic vector. Thus, if the fields 
$(X,P)$ may be thought of as describing a map into complexified cotangent bundle $T^*M$ of complexified space--time, imposing this constraint and gauge symmetry mean that the target space of~\eqref{boson-str} is the space of complex null geodesics. Note that, unlike the particle case, $P_\sigma$ is only defined up to a rescaling ($P$ takes values in the canonical bundle of $\Sigma$) so there is no preferred scaling of these geodesics.

In four dimensions, this space of complex null geodesics lies in the product of twistor space and its dual and so has become known as (projective) ambitwistor space, denoted  
$\PA$. It was studied in the 1970s and 1980s as a vehicle for extending the deformed twistor space constructions for Yang-Mills \cite{Witten:1978xx,Isenberg:1978kk}.  Such 
constructions were extended to arbitrary dimensions in the context of gravity by LeBrun \cite{LeBrun:1983} and in a supersymmetric context in 10 dimensions by 
Witten \cite{Witten:1985nt}.  
See
also~\cite{Mason:2005kn,Beisert:2012xx} for more recent work on
ambitwistors in the context of scattering amplitudes in $\cN=4$ super
Yang-Mills. 
Although the connection between spaces of complex null geodesics with twistors is less direct in higher dimensions, we will use the term `ambitwistor 
space' throughout as they nevertheless provide a family of twistor--like correspondences that encode space--time fields into holomorphic objects on the space of (perhaps 
spinning) complexified null geodesics in arbitrary dimensions. In particular, as in the usual twistor correspondence, deformations of the space--time metric may be encoded in deformations of the complex structure of ambitwistor space. Similar to the original twistor string, the fact that these ambitwistor string theories are chiral (holomorphic) allows them to describe space--time gravity by coupling to the complex structure of the target space, here $\PA$.  We will see that  the integrated vertex operators for the ambitwistor string describe deformations of the complex structure of $\PA$ preserving this contact structure and naturally incorporate delta function support on the scattering equations~\eqref{scatt}.  Indeed these are necessary to impose the resulting constraint $P^2=0$ everywhere on $\Sigma$ which is crucial to reduce the target space from $T^*M$ to $\PA$.

\medskip

Since the spectrum of this string theory contains only massless states, and since the constraint $P^2=0$ that reduced the target space from $T^*M$ to $\PA$ is the same constraint as results from imposing the scattering equations~\eqref{scatt}, one might expect this model to underpin the formulae for scattering massless particles of spin $s=0,1,2$ found in~\cite{Cachazo:2013hca,Cachazo:2013iea}. This turns out to be essentially correct for the spin zero case (after coupling to a worldsheet current algebra). To recover the S--matrices of Yang-Mills and gravity, we must instead 
start from the worldline action 
\be\label{spin-null-ptcle}
	S[X,P,\Psi]=\int  P_\mu \rd X^\mu+ g_{\mu\nu}\Psi^\mu\rd\Psi^\nu -  \frac{e}{2}P_\mu P^\mu - \chi P_\mu\Psi^\mu
\ee
describing a massless particle with spin. Here, $\Psi^\mu$ is a wordline fermion and $\chi$ imposes a constraint associated to the worldline supersymmetry acting as
\be
	\delta X^\mu=\epsilon \Psi^\mu\qquad  \delta\Psi^\mu=\epsilon P^\mu\qquad\delta P_\mu=0
\ee
on the matter fields and
\be	
	 \delta \chi= \rd\epsilon\qquad\qquad \delta e = \epsilon \chi
\ee
on the gauge fields. The space of solutions to the field equations modulo these gauge transformation is the space of (parametrized) spinning null geodesics. Quantization of 
$\Psi^\mu$ gives the Dirac matrices and the quantization of the constraint $\Psi^\mu P_\mu=0$ is the massless Dirac equation.

In section~\ref{sec:super-ambitwistors} we consider a chiral analogue of the spinning ambitwistor string with worldsheet action
\be\label{spin-string}
	S[X,P,\Psi]=\int_\Sigma   P_\mu\delbar X^\mu+\frac{e}{2} P_\mu P^\mu +\sum_{r=1}^2 \Psi_{r\mu} \delbar\Psi_r^\nu + \chi_r P_\mu \Psi_r^\mu
\ee
with two spin vectors $\Psi_r^\mu$ each of which also transforms as worldsheet spinor (so that each $\Psi = \Psi_\sigma \sqrt{\rd\sigma}$ in local coordinates). We will call these theories `type II ambitwistor strings'. Note that here, in stark contrast to the usual RNS string, {\it both sets of $\Psi_r$ fields are left-moving}. The path integral over these fermions leads to the Pfaffians in the representation of the tree--level gravitational S--matrix found by Cachazo {\it et al.}. As we show in section~\ref{sec:heterotic}, trading one set of these fermions for a general current algebra as in the heterotic string gives (at leading trace) their representation of Yang-Mills amplitudes where one Pfaffian is replaced by a current correlator. Trading both sets of fermions for general current algebras replaces both Pfaffians by current correlators, giving the amplitudes for scalars in the adjoint of $G\times\widetilde G$ found in~\cite{Cachazo:2013iea}. Thus the origin of  `gravity as Yang-Mills squared' in~\cite{Cachazo:2013iea} is really the same as in the original KLT construction~\cite{Kawai:1985xq}.

We conclude in section~\ref{sec:conclusions} with a brief look at some of the many possible directions for future work and new perspectives offered by these ideas. 
These include a brief look at the Ramond-NS and Ramond-Ramond sectors where we anticipate space--time spinors and form fields to reside, and a discussion of how to extend these amplitudes and the scattering equations to higher genus. In section~\ref{other} we briefly explain how to define Green-Schwarz ambitwistor string actions that make direct contact with Witten's super ambitwistor space~\cite{Witten:1985nt} for 10 dimensional space--time.  It also seems likely that there is a pure spinor formulation.   In section~\ref{sec:twistor-strings} we argue that the existing twistor string models are perhaps best thought of as different representations of these theories. 

These ideas should also lead to new insights into the BCJ colour
kinematics relations.  Although these have their origins in standard
string theory, see {\it e.g.}~\cite{BjerrumBohr:2010hn}, ambitwistor
strings give a simpler context without the towers of massive modes of
standard string theory.  Ambitwistors may also provide a route towards
a conventional field theory formulation of these ideas, perhaps using
the scattering equations as in {\it e.g.}~\cite{Monteiro:2013rya}, or
an ambitwistor action such as in~\cite{Mason:2005kn}.


\section{The space of complex null geodesics}
\label{sec:brief-rev}

The target space of the string theories we construct will be the space of complex null geodesics in complexified space--time $M$. We denote the space of {\it scaled} complex null geodesics by $\A$ and the space of {\it un}scaled complex null geodesics by $\PA$, calling them `ambitwistor space' and `projective ambitwistor space', respectively. The terminology follows the four dimensional case where $\PA$ can be viewed as the projectivized cotangent bundle of both the twistor and dual twistor spaces of $M$\footnote{In fact, in four dimensions, $\PA$ sits as a quadric inside the Cartesian product of twistor space and its dual; see~\ref{sec:twistor-strings}.}. However, ambitwistor space is a more versatile notion that exists in any dimension and for any (globally hyperbolic) space--time. It has long been known that gauge and gravitational fields may be encoded in terms of holomorphic structures on $\PA$~\cite{Witten:1978xx,Isenberg:1978kk,LeBrun:1983}.  We will discuss the gauge theory case later, but here give a brief review of the gravitational case following LeBrun~\cite{LeBrun:1983} (see also appendix~\ref{ambi-app}).

Given any $d$ dimensional space--time $(M_\R,g_\R)$, its complexification $(M,g)$ is a Riemannian manifold of complex dimension $d$ with a holomorphic metric $g$. A complex null direction at a point $x\in M$ is a tangent vector $v\in T_xM$ obeying $g(v,v)=0$, or equivalently a cotangent vector $p\in T^*_xM$ obeying $g^{-1}(p,p)=0$. The bundle $T^*_NM$ of complex null directions over $M$ thus sits inside the holomorphic cotangent bundle $T^*M$ as
\be\lab{TstarNM}
	T^*_NM = \left\{ (x,p)\in T^*M\, | \, g^{-1}(p,p) = 0\right\}
\ee
To obtain the space $\A$ of scaled complex null geodesics, we must quotient $T^*_NM$ by the action of 
\be
	D_0=p^\mu\left(\frac{\del}{\del x^\mu} + \Gamma_{\mu\nu}^\rho\, p_\rho \frac{\del}{\del p_\nu}\right)\,.
\label{spray}
\ee
This vector is the horizontal lift of the space--time derivative $p^\mu\del_\mu$ to the cotangent bundle $T^*M$ using the Levi-Civita connection $\Gamma$ associated to $g$. Flowing along $D_0$ generates a null geodesic -- the integral curves of $D_0$ are the horizontal lifts of geodesics with (null) cotangent vector $p_\mu$ to the cotangent bundle $T^*M$ -- so to obtain $\A$ we should not count as different two points in $T^*_NM$ that are joined along this flow. 

Ambitwistor space is a holomorphic symplectic manifold. To see this, note that the cotangent bundle $T^*M$ is naturally a holomorphic symplectic manifold with holomorphic symplectic form $\omega=\rd p_\mu\wedge\rd x^\mu$. The geodesic spray $D_0$ of~\eqref{spray} is the simply the Hamiltonian vector field associated to the function $\frac{1}{2}g^{\mu\nu}(x)p_\mu p_\nu$; that is,
\be
	D_0\hook\omega + \frac{1}{2} \rd (p^\mu p_\mu) = 0\,.
\ee
Thus, to both impose the constraint $p^2=0$ and quotient by the action of $D_0$ is simply to take the symplectic quotient of $T^*M$ by $D_0$, and so $\A$ naturally inherits a holomorphic symplectic structure. As $\cL_{D_0}\omega=0$, the symplectic form is invariant along these null geodesics  and we will abuse notation by also using $\omega$ to denote the holomorphic symplectic form on $\A$. For a $d$ dimensional space--time, $\A$ is $2d-2$ (complex) dimensional and the fact that the symplectic structure is non--degenerate means that $\omega^{d-1}\neq0$.

The null geodesics obtained this way come with a natural scaling that may be adjusted by rescaling $p \to r p$ for any  non--zero complex number $r$. On $T^*M$, this scaling is generated by the Euler vector field $\Upsilon = p_\mu\del / \del p_\mu$ and, since $[\Upsilon,D_0]=D_0$, the scaling descends to $\A$. If we further quotient $\A$ by the action of $\Upsilon$ we obtain the $2d-3$ (complex) dimensional space $\PA$ of unscaled complex null geodesics.

To understand the geometric structure inherited by $\PA$, note that the natural symplectic potential $\theta = \Upsilon\hook\omega=p_\mu\rd x^\mu$ on $T^*M$ obeys $\cL_{D_0}\theta + \frac{1}{2}\rd (p_\mu p^\mu)=0$. Thus, while $\theta$ is not invariant along the flow of an arbitrary geodesic, it is invariant along (lifts to $T^*M$ of) null geodesics and so descends to $\A$. The projectivization $\A\rightarrow P\A$ expresses  $\A$ as the total space of a line bundle that we denote $L^{-1}\rightarrow P\A$; sections of $L$ are functions of homogeneity degree one in $p$.  Finally, since $\cL_\Upsilon\theta =\theta$, the symplectic potential $\theta$ descends to the $2d-3$ dimensional manifold $P\A$ to define a 1-form with values in  $L$, $\theta\in\Omega^1(\PA,L)$.  Such a line bundle--valued 1-form is known as a contact structure. Because the symplectic structure $\omega$ on $\A$ obeys $\omega^{d-1}\neq 0$, the contact 1-form $\theta$ on $\PA$ obeys $\theta\wedge\rd\theta^{d-2} \neq 0$ and is said to be non--degenerate. Thus,  a $d$ dimensional complex space--time $(M,g)$ has a space of complex null geodesics $\PA$ that is a $2d-3$ dimensional complex non--degenerate contact manifold.

While a point of $\PA$ by definition corresponds to a complex null geodesic in $M$,  a point in $M$ corresponds to a quadric surface $Q_x\subset\PA$. This may be viewed as the space of complex null rays through $x$. For example, in four dimensions $Q_x\cong\CP^1\times\CP^1$ parametrizing the complex null vectors $p_{\alpha\dot\alpha}= \lambda_\alpha\tilde\lambda_{\dot\alpha}$ up to scale. For the real Minkowski slice, we set $\tilde\lambda_{\dot\alpha} = (\lambda_\alpha)^*$ which gives the familiar celestial sphere $S^2\subset\CP^1\times\CP^1$. More generally, the correspondences between space--time $M$ and the space of complex null geodesics with or without scaling may be expressed in terms of double fibrations as 
\be\lab{doublefibration}
\begin{aligned}
\begin{picture}(50,40)
\put(0.0,0.0){\makebox(0,0)[c]{$\A$}}
\put(60.0,0.0){\makebox(0,0)[c]{$M$}}
\put(34.0,33.0){\makebox(0,0)[c]{$T^*_NM$}}
\put(5.0,18.0){\makebox(0,0)[c]{$\pi_1$}}
\put(55.0,18.0){\makebox(0,0)[c]{$\pi_2$}}
\put(25.0,25.0){\vector(-1,-1){18}}
\put(37.0,25.0){\vector(1,-1){18}}
\end{picture}\hspace{3cm}
\begin{picture}(50,40)
\put(0.0,0.0){\makebox(0,0)[c]{$\PA$}}
\put(60.0,0.0){\makebox(0,0)[c]{$M$}}
\put(34.0,33.0){\makebox(0,0)[c]{$PT^*_NM$}}
\put(5.0,18.0){\makebox(0,0)[c]{$\pi_1$}}
\put(55.0,18.0){\makebox(0,0)[c]{$\pi_2$}}
\put(25.0,25.0){\vector(-1,-1){18}}
\put(37.0,25.0){\vector(1,-1){18}}
\end{picture}
\end{aligned}
\ee
where, in the projective case the fibres of $\pi_2$ are the unscaled complex lightcones $Q_x$ and are compact holomorphic quadrics of complex dimension $d-2$, while the fibres of $\pi_1$ are the complex null geodesics.

\medskip

LeBrun~\cite{LeBrun:1983} shows that, conversely, $\PA$ together with its contact structure on is sufficient to reconstruct the original space--time $M$, together with its torsion--free conformal structure. In outline, to reconstruct $M$ from $\PA$ one first notes that the non--degenerate contact structure $\theta$ defines a complex structure on $\PA$. To see this, we use the fact that because $\theta$ is non--degenerate, $\theta\wedge \rd\theta^{d-2}$ is a non--vanishing $2d-3$ form on the $2d-3$ complex dimensional space. We then simply declare an antiholomorphic vector to be a vector $\overline V$ which obeys $\overline V\hook(\theta \wedge \rd\theta^{d-2})=0$. Now, supposing we can find at least one holomorphic quadric $Q_0\subset\PA$ with normal bundle $T\bP_{d-1}\otimes\cO(-1)|_{Q_0}$, Kodaira theory assures us that we can find a $d$ dimensional family of nearby $Q_x$ (see {\it e.g.}~\cite{LeBrun:1983} for details). We then interpret this family as providing the points in space--time $M$. The conformal structure on $M$ together with its null geodesics may be reconstructed from the intersection of these $Q_x$ in $\PA$. LeBrun shows~\cite{LeBrun:1983} that these geodesics arise from a torsion--free connection precisely when $P\A$ admits a contact structure $\theta$ that vanishes on restriction to the $Q_x$. Furthermore, arbitrary small deformations of the complex structure of $\PA$ which preserve the contact structure $\theta$ correspond to small deformations of the conformal structure on $M$.

\medskip

We will use a linearized version of this correspondence in order to generate amplitudes, focussing on the gravitational case.  See appendix~\ref{ambi-app} or~\cite{Baston:1987av} for a more detailed discussion of  the linear Penrose transform for the ambitwistor correspondence in the case of general spin.  Since the conformal structure of $M$ is determined by the contact structure of $\PA$, to describe a fluctuation in the space--time metric we need only consider a perturbation $\delta\theta$ of the contact structure.   Up to infinitesimal diffeomorphisms, $\delta\theta$ can be taken to be an antiholomorphic 1-form with values in the contact line bundle. If $\delta\theta$ is $\delbar$-exact then it does not genuinely describe a deformation of the contact structure, but rather just a diffeomorphism of $\PA$ along a Hamiltonian vector field. Thus non--trivial deformations correspond to elements of the Dolbeault cohomology class $[\delta\theta]$. In short,
\be\label{eta}
	\delta \theta \in \Omega^{0,1}(L)  \, , \qquad [\delta\theta ] \in H^{0,1}(\PA,L)\, .
\ee
Pulled back to the non--projective space $\A$, it determines a $(0,1)$-form valued Hamiltonian vector field $X_{\delta\theta}$ by
\be
	X_{\delta\theta}\hook \omega + \rd (\delta\theta)=0\,, \qquad X_{\delta\theta}\in H^{0,1}(\A,T_\PA)
\ee
and so $X_{\delta\theta}$ determines a deformation of the complex structure of $\A$ and hence $P\A$. To see how this deformation determines a deformation of the conformal structure on $M$, we first pull it back by $\pi_1$ to obtain $\pi_1^*(\delta\theta)$ on $PT^*_NM$.   It turns out that there is no first cohomology on $PT^*_NM$ because as a complex manifold it is essentially the cartesian product of $M$, which has no cohomology by assumption, and a projective quadric of dimension $d-2$, which has no first cohomology in dimension $d>3$, and none with this weight for any $d$.  Thus we can write
\be 
	\pi_1^*(\delta\theta)=\delbar j
\ee
for some $j\in \Gamma (PT^*_NM,L)$. Now, because $\delta\theta$ was originally defined on $\PA$, its pullback to $PT^*_NM$ must be constant along the fibres of $\pi_1$ and so  $D_0(\pi_1^*(\delta\theta))=0$. But because $[D_0,\delbar]=0$ as $D_0$ is a holomorphic vector field, we learn that $\delbar (D_0j)=0$, or in other words that $D_0j$ is holomorphic. Finally, because $D_0j$ is homogeneous of degree 2 in $p_\mu$ and holomorphic, it must actually be quadratic so that
\be
	h:=D_0j= \delta g^{\mu\nu}(x)\,p_\mu p_\nu
\ee
for some symmetric, trace--free tensor $\epsilon^{\mu\nu}(x)$ depending only on $x$.  $\delta g^{\mu\nu}$ describes a variation in the space--time metric, while $h$ itself can be viewed as the deformation of the Hamiltonian constraint $g^{\mu\nu}p_\mu p_\nu=0$. To summarize, the ambitwistor Penrose transform relates deformations of the conformal structure on space--time to elements of $H^{0,1}(\PA,L)$ on projective ambitwistor space. The case of particles with more general spin is treated in appendix~\ref{ambi-app} following~\cite{Baston:1987av}.

One of the most important differences between this ambitwistor version of the Penrose transform and the (perhaps more familiar) Penrose transform between twistor space and 
space--time is that here, the field on space--time is not required to satisfy any field equations at this stage. Much work in the 70's and 80's focussed on the expression of the field 
equations in ambitwistor space (in terms of the existence of supersymmetries~\cite{Witten:1978xx,Witten:1985nt} or (essentially equivalently) formal 
neighbourhoods~\cite{Isenberg:1978kk, Baston:1987av, LeBrun:1991jh}).  In the following we will see that for our string models, the space--time massless field equations arise 
automatically from quantum consistency of the symplectic reduction at the level of the worldsheet path integral.

The key example that we will use to discuss scattering amplitudes is the case where our metric fluctations correspond to momentum eigenstates in flat space.  To describe these space--time momentum eigenstates in terms of wavefunctions on ambitwistor space we take $\delta g^{\mu\nu}(x) = \epsilon^{\mu\nu} \e^{\im k\cdot x}$ whereupon $h$ becomes
\be
	h=\e^{\im k\cdot x}\epsilon^{\mu\nu}p_\mu p_\nu\,
\ee
while $j= D_0^{-1}h$ and $\delta\theta$ are then given by
\be
\label{eta-k-estate}
	j= \frac {\e^{\im k\cdot x}\epsilon^{\mu\nu}p_\mu p_\nu}{k\cdot p}\, , 
	\qquad \delta\theta= \bar \delta(k\cdot p)\, \e^{\im k\cdot x}\epsilon^{\mu\nu}p_\mu p_\nu\, .
\ee
As promised, $\delta\theta$ is a (0,1)-form on $\PA$ of homogeneity $+1$ in $p$, and so defines an element of $H^{0,1}(\PA,L)$. 

The form of the ambitwistor wavefunction $\delta\theta$ is somewhat similar to the form $\sim \bar\delta\left(\langle \lambda\, \lambda_i\rangle\right) \e^{\im [\mu,\tilde\lambda_i]}$ of a twistor wavefunction for a four--dimensional momentum eigenstate with four--dimensional momentum $k = \lambda_i\tilde\lambda_i$. The main differences are that {\it i)} the ambitwistor wavefunction is non--chiral and is defined in arbitrary dimensions, and {\it ii)} neither the momentum nor the (symmetric, trace--free) polarization vector are constrained in the ambitwistor wavefunction. In particular, at this stage we do note require $k^2=0$ or $k_\mu\epsilon^{\mu\nu}=0$. This is in keeping with the fact that holomorphic objects on ambitwistor space are not manifestly on--shell objects in space--time. As mentioned above, these constraints will arise from quantum consistency of the string theory, but it is worth noting that the formulae of~\cite{Cachazo:2013hca,Cachazo:2013iea} involve polarization vectors $\epsilon^{\mu\nu}$ and momenta $k$ --- their representation of amplitudes is also not manifestly on--shell. Finally, we remark that in the context of the ambitwistor string path integral, the factor of $\bar\delta(k\cdot p)$ in the ambitwistor wavefunction for a momentum eigenstate ultimately provides the origin of the constraint to solutions of the scattering equations in the formulae of~\cite{Cachazo:2013hca,Cachazo:2013iea}.


\section{The bosonic ambitwistor string}
\label{sec:boson}

We now consider a chiral string theory whose target space is projective ambitwistor space. As discussed in the introduction, the worldsheet action is a natural analogue of the worldline action for a massless scalar particle and may be written as
\be\label{bos-action}
	S_{\rm bos}=\frac1{2\pi}\int_\Sigma P_\mu \delbar X^\mu- \frac{e}{2} P_\mu P^\mu\  .
\ee
Note that this is different from the first--order action 
\be
	S' = \frac{1}{2\pi}\int_\Sigma P_\mu \rd X^\mu  - \frac{1}{2} P_\mu \wedge *P^\mu
\lab{Polyakov}
\ee
that is equivalent to the usual Polyakov string, because in~\eqref{Polyakov} $P_\mu$ is a general 1-form on the worldsheet, {\it i.e.} $P_\mu\in \Omega^1 \cong\Omega^{1,0}\oplus\Omega^{0,1}$, whereas in~\eqref{bos-action}  $P_\mu$ lives only in $\Omega^{1,0}\cong K$ and the kinetic operator is $\delbar$ rather than the full exterior derivative. We interpret $P^2$ in~\eqref{bos-action} to be a quadratic differential and then $e\in\Omega^{0,1}(T_\Sigma)$ is a Beltrami differential.

Both~\eqref{bos-action} and~\eqref{Polyakov} are manifestly invariant under worldsheet reparametrizations. In particular, under a diffeomorphism generated by a smooth worldsheet vector field $v\in T_\Sigma$, the fields in~\eqref{bos-action} transform as
\be
	\delta X^\mu = v\del X^\mu\,,\qquad \delta P_\mu =\del(vP_\mu) \,,\qquad \delta e = v\del e-e\del v 
\ee
as usual. However, $S_{\rm bos}$ is also separately invariant under the gauge transformations
\be\label{symmetry}
	\delta X^\mu = \alpha P^\mu\,,\qquad \delta P_\mu = 0\,,\qquad \delta {e}= \delbar \alpha 
\ee
for $\alpha$ a further smooth worldsheet vector. As explained in section~\ref{sec:brief-rev}, together with the associated constraint $P^2=0$, these gauge transformations implement the symplectic reduction from $T^*M$ to the space of (scaled) null geodesics $\A$. Furthermore, since $P$ takes values in the line bundle $K$, it is only defined up to a local rescaling. Thus there is really no preferred scaling so the target space is properly interpreted as $\PA$. Said differently, we are identifying the pullback of the contact line bundle $L$ with $K$, and then the worldsheet action is simply the pullback to $\Sigma$ of the contact 1-form $\theta$ on $\PA$.

\subsection{The BRST operator}

To perform these gauge redundancies in the quantum theory, we introduce the usual holomorphic reparametrization ghost $c$ and antighost $b$, which are fermionic sections of $T_\Sigma$ and $K^2$, respectively. In addition, we introduce a further set of ghosts and antighosts associated to the gauge symmetry~\eqref{symmetry}. We call these new ghosts $\tilde c$ and $\tilde b$, and they are again fermionic sections of $T_\Sigma$ and $K^2$ -- that is, despite the tildes, they are again holomorphic on the worldsheet. The fact that we have two sets of the usual holomorphic ghosts but no antiholomorphic ghosts is in keeping with the chiral nature of the model. It will have consequences for the form of the vertex operators that we explore below.

At genus zero $h^1(\Sigma,T_\Sigma)=0$ so we can use the gauge symmetry $\delta e = \delbar\alpha$ to set $e=0$. In this gauge, the ghost action takes the standard form
\be
	S=\frac{1}{2\pi}\int_\Sigma b \delbar c+ \tilde b\delbar \tilde c
\ee
while the BRST operator is
\be\label{Q-bos}
	Q=\oint c T+ \frac{\tilde c}{2} P^2
\ee
where the worldsheet stress tensor $T=P_\mu \del X^\mu + c\del b + 2(\del c)b + \tilde b\del \tilde c$.  The central charge is 
\be
	{\rm c}=2d-26-26=2(d-26)\, .
\ee
Thus $Q^2=0$ when $d=26$ as in the standard bosonic string. However, here we recall that $X$ defines a map into the complexification of space--time.

\subsection{Vertex operators}

As in section~\ref{sec:brief-rev}, the simplest vertex operators correspond to variations in the space--time metric $g\to g+ \delta g$, where for momentum eigenstates $\delta g^{\mu\nu}(X)= \epsilon^{\mu\nu}\e^{\im k\cdot X}$ with $\epsilon^{\mu\nu}$ symmetric and trace--free. The corresponding fixed vertex operators are
\be\label{bos-fixed}
	c\tilde c V := c \tilde c\, P_\mu P_\nu\epsilon^{\mu\nu} \e^{\im k\cdot X}\, ,
\ee
and may be interpreted as $c\tilde c$ times the variation in $P^2$ under this variation of the space--time metric.  Note that the quadratic differential $P_\mu P_\nu \epsilon^{\mu\nu}\e^{\im k\cdot X}$ is balanced by the ghosts $c, \tilde c\in T_\Sigma$ to form a scalar operator, and that the trace  $\epsilon^\mu_{\ \mu}$ is absent because we enforce $P^2=0$. This vertex operator is BRST closed iff the momentum and polarization obey
\be\lab{on-shell}
	k^2=0\, , \quad \epsilon^{\mu\nu} k_\mu=0
\ee
where, as usual, these conditions come from double contractions with the BRST operator. Similarly, it is BRST exact if $\epsilon^{\mu\nu}=k^{(\mu}\epsilon^{\nu)}$ for some $\epsilon^\nu$, which is usual linearized diffeomorphism invariance. Consequently, the vertex operator~\eqref{Vboson} represents an on--shell linearized graviton.

The corresponding integrated vertex operators take the form
\be\lab{Vboson}
	\int_\Sigma\cV :=\int_\Sigma \bar\delta(k\cdot P)\,V=\int_\Sigma \bar\delta (k\cdot P)\,P_\mu P_\nu\, \epsilon^{\mu\nu}\e^{\im k\cdot X}\ .
\ee
The fact that we remove the ghost $c$ from the fixed vertex operator is standard, but the presence of the $\bar\delta(k\cdot P)$ here appears to be non--standard and requires further explanation. Firstly, notice that $\cV$ is indeed a (1,1)-form on the Riemann surface so that~\eqref{Vboson} is at least well-defined. As usual, $\int_\Sigma \cV$ may be interpreted as a deformation of the worldsheet action induced by the deformation $\delta g$ of the space--time metric. To understand this, recall that our worldsheet action is really just the pullback to $\Sigma$ of the contact 1-form $\theta$ on $\PA$, where the pullback to $\Sigma$ of the contact line bundle $L\to\PA$ is identified with the worldsheet canonical line bundle $K$. From the discussion of section~\ref{sec:brief-rev} we know that a variation of the space--time metric $\delta g$ determines and is determined by a deformation of this contact 1-form $\theta\to\theta+\delta\theta$ where $\delta\theta$ defines a class  $[\delta\theta]\in H^{0,1}(\PA,L)$. Pulled back to the worldsheet, $[\delta\theta]$ thus lies in $H^{0,1}(\Sigma,K)$ and may be integrated to produce a deformation of the action. The vertex operator \eqref{Vboson} is just this deformation specified to the case of a momentum eigenstate~\eqref{eta-k-estate} on ambitwistor space. From this point of view, the fixed vertex operator is the Hamiltonian associated to the reduction from $PT^*_NM$ to $\PA$. Again, the vertex operator~\eqref{Vboson} is BRST closed iff the on--shell conditions~\eqref{on-shell} hold. Thus the field equations are not automatically built into the ambitwistor correspondence, but arise in the usual manner through quantum consistency of the string model.

\medskip

Perhaps the most important difference between the ambitwistor string~\eqref{bos-action} and the usual string is that the $XX$ OPE in~\eqref{bos-action} is trivial. (This is simplest to see in the gauge $e=0$).  In particular, $\e^{\im k\cdot X}$ does not here acquire anomalous conformal weight, so we cannot compensate for the conformal weight of a generic polynomial in $P$ or $\del^rX$ by allowing $k^2\neq0$. Consequently, there are no massive states in the spectrum, which is consistent with the ambitwistor string being a chiral $\alpha'\to0$ limit of the usual string (see appendix~\ref{sec:alpha0}).


\subsection{The path integral and the scattering equations}

At genus zero, the three zero--modes for each of $c$ and $\tilde c$ require that we insert three fixed vertex operators~\eqref{bos-fixed} and then arbitrarily many integrated ones~\eqref{Vboson}.  Thus the $n$-particle amplitude is given by the worldsheet correlation function 
\be\lab{bos-pi}
	\cM(1,\ldots,n)= \left\la c_1\tilde c_1V_1\, c_2\tilde c_2 V_2\, c_3\tilde c_3 V_3\int \cV_4\cdots\int \cV_n \right\ra\, .
\ee
Consider first the $XP$ system. The vertex operators are polynomial neither in $P$ nor in $X$, so to evaluate this correlation function it is simplest to incorporate the plane waves $\e^{\im k_i\cdot X}$ into the action. In the gauge $e=0$ this becomes
\be
	S[X,P]=\frac1{2\pi}\int_\Sigma P_\mu\delbar X^\mu + \im \sum_{i=1}^n k_i\cdot X\, \delta^2(\sigma-\sigma_i)
\ee
and now contains the entire $X$ dependence inside the path integral.  Let us consider integrating out $X$. The constant zero modes decouple from the kinetic $P\delbar X$, so integrating these out leads to a momentum conserving $\delta$-function $\delta^{26}(\sum k_i)$ as usual. The non--zero modes are Lagrange multipliers enforcing the field equation
\be
	\delbar P_\mu= 2\pi \im \sum_i k_{i\mu} \delta^2(\sigma-\sigma_i)
\ee
on the worldsheet (1,0)-form $P_\mu$. At genus zero, this has unique solution
\be\lab{Pconst}
	P_\mu(\sigma)=\rd \sigma\sum_{i=1}^n \frac {k_{i\mu} }{\sigma-\sigma_i}\ ,
\ee
which may now be substituted into the remaining factors of $P_\mu$ in the vertex operators. In particular, using the on--shell conditions $k_i^2=0$,  the factors of $\bar\delta(k_i\cdot P(\sigma_i))$ impose the scattering equations
\be
	\sum_{j\neq i}\frac {k_i\cdot k_j}{\sigma_i-\sigma_j}=0
\ee
of Gross \& Mende~\cite{Gross:1987ar,Gross:1987kza}, which are sufficient to determine the insertion points $\sigma_i$ in terms of the external momenta. However, unlike the saddle--point approximation used in~\cite{Gross:1987ar,Gross:1987kza}, here these scattering equations provide the only contributions to the path integral without taking any kinematic limit. This is the same situation as found in the expressions for massless amplitudes found in~\cite{Cachazo:2013hca,Cachazo:2013iea} and is also the same as in the twistor string in four dimensions~\cite{Witten:2004cp}.

Just like the $c$ ghosts, the zero modes of $\tilde c$ give a factor of $(\sigma_{12}\sigma_{23}\sigma_{13}) / (\rd \sigma_1 \rd\sigma_2 \rd\sigma_3)$.  Including this contribution, the measure
\be
	{\prod_i}' \bar\delta(k_i\cdot P(\sigma_i))  := \frac{\sigma_{12}\sigma_{23}\sigma_{13}}{\rd\sigma_1\rd\sigma_2\rd\sigma_3} \prod_{i=4}^n \bar \delta (k_i\cdot P(\sigma_i))
\ee
transforms under M{\" o}bius transformations as worldsheet vector at each point, and was shown in~\cite{Cachazo:2013gna} to be permutation invariant (on the support of the 
overall momentum conserving $\delta$-function). Thus we the path integral~\eqref{bos-pi} gives
\be
	\cM(1,\ldots,n)=\delta^{26}\left(\sum_ik_i\right)\int \frac{1}{\rm Vol\,SL(2;\C)}\ {\prod_i}' \,\bar\delta (k_i\cdot P(\sigma_i))\ 
	\prod_{j=1}^n \epsilon_j^{\mu\nu}P_{\mu}(\sigma_j)P_\nu(\sigma_j)\, ,
\ee
where $P_\mu(\sigma)$ is constrained to take its value as in~\eqref{Pconst} and where the factor of $1/{\rm Vol\,SL(2;\C) = (\sigma_{12}\sigma_{23}\sigma_{31})/(\rd\sigma_1\rd\sigma_2\rd\sigma_3)}$ is the usual $c$ ghost path integral. Unfortunately we do not have a satisfactory interpretation of these amplitudes in relation to a standard space--time theory of gravity\footnote{Their three particle amplitudes are suggestive of a (Weyl)$^3$ vertex, while the overall weights in the momenta seem to extend these vertices to $n$-point amplitudes using a standard $1/k^2$ propagator.}.   In section~\ref{sec:super-ambitwistors} we turn to a chiral analogue of a type II RNS string model, which does yield the correct gravitational amplitudes. We return to consider this bosonic model in section~\ref{sec:heterotic} where we will see that, after including two worldsheet current algebras, it does provide the correct amplitudes in a certain  scalar theory.


\section{Ambitwistor superstrings}
\label{sec:super-ambitwistors}

In this section we construct the worldsheet theory underlying the representations of gravitational scattering amplitudes found in~\cite{Cachazo:2013hca,Cachazo:2013iea}. As mentioned in the introduction, our starting--point is a chiral worldsheet analogue of the wordline action for a massless spinning particle. Thus, in addition to the $(P,X)$ system above, we choose a spin structure $\sqrt K $ on $\Sigma$ and introduce two additional fermionic fields $\Psi^\mu _r$ ($r=1,2$), each with values in $\sqrt K \otimes X^*TM$. Furthermore, as well as gauging $P^2$, we will also gauge the $\Psi_r\cdot P$ with analogues of worldsheet gravitini $ \chi_r \in \Omega^{0,1}\otimes \sqrt {T_\Sigma}$.  These constraints will have the interpretation of reducing the target space of the model to super ambitwistor space, as we discuss in section~\ref{sec:superambi}.

The action of the matter fields is taken to be
\be\label{fermions}
	S_f=\frac1{2\pi}\int_\Sigma P_\mu\delbar X^\mu - \frac{e}{2} P^2 + \sum_{r=1,2}\frac{1}{2}\Psi_{r\mu}\delbar \Psi^{\mu}_r  - \chi _r P_\mu\Psi_r^\mu\, ,
\ee
In addition to the transformations
\be\label{extended}
	\delta X^\mu = \alpha P_\mu\,,\qquad\delta\Psi^\mu = 0\,,\qquad \delta P_\mu = 0\,,\qquad\delta e = \delbar \alpha\,,\qquad \delta\chi_r = 0
\ee
that trivially extend~\eqref{symmetry}, this action also has a degenerate $\cN=2$ worldsheet supersymmetry generated by
\be\label{susy}
	\delta X^\mu=\epsilon_r \Psi_r^\mu\, , \qquad \delta \Psi_r^\mu= \epsilon_r P^\mu\, , \qquad\delta  P_\mu=0 \, , \qquad \delta e =0\,,\qquad\delta \chi _r=\delbar \epsilon_r \, ,
\ee
where $\epsilon_r\in T^{1/2}$ are a pair of anticommuting worldsheet spinors. We will discuss the meaning of this gauge symmetry presently, but first note that there is also a $\Z_2\times\Z_2$ symmetry acting as $\Psi_r \to -\Psi_r$ and $\chi_r \to-\chi_r$ independently on each set of fermion species $r$. We will gauge this discrete symmetry, meaning we only consider vertex operators that are invariant under $\Z_2\times\Z_2$.  In particular, requiring invariance under the action of this $\Z_2\times\Z_2$ means we break the $O(2)$ symmetry of~\eqref{fermions} that rotates the two fermion species into one another down to the $\Z_2\subset O(2)$ that simply exchanges them.

\subsection{The super ambitwistor correspondence}
\label{sec:superambi}

The underlying geometry of this string leads to an extension\footnote{This RNS--type extension is somewhat different to the notion of superambitwistor space used in~\cite{Witten:1978xx,Witten:1985nt} where space--time supersymmetry is manifest. See section~\ref{other} for a brief discussion of a Green-Schwarz ambitwistor string.} of the bosonic ambitwistor correspondence that was described in the section~\ref{sec:brief-rev}.  The fields $(X^\mu,P_\mu,\Psi_r^\mu)$ define a map from the worldsheet into the bundle $T^*_SM:=(T^*\oplus \Pi T\oplus\Pi T)M$, where the $\Pi$ reminds us that the two tangent vectors $\Psi^\mu_r$ are each anticommuting. We let $(x^\mu,p_\mu,\psi_r^\mu)$ denote coordinates on this space. $T^*_SM$ is naturally a holomorphic symplectic supermanifold with holomorphic symplectic potential
\be\label{ssymp}
	\theta_S = p_\mu\rd x^\mu + \sum_{r=1}^2\frac{1}{2}g_{\mu\nu}(x) \psi^\mu_r \rd\psi^\nu_r
\ee
and associated symplectic form $\omega_S = \rd\theta_S$. Note that the fermionic differential 1-forms $\rd \psi$ are commuting. Imposing the constraints $p^2=0$ and $p_\mu \psi_r^\mu=0$ gives what we will call the bundle of super null covectors $T^*_{SN}M$, {\it i.e.},
\be
	T^*_{SN}M:=\left\{ (x^\mu,p_\mu ,\psi^\mu_r)\in (T^*\oplus \Pi  T\oplus\Pi T)M\ |\ p^2=0=p_\mu\psi_r^\mu\right\}\, .
\ee
As before, with the help of the symplectic form $\omega_S$, the functions $\frac{1}{2}p^2$ and $p_\mu\psi_r^\mu$ define Hamiltonian vector fields $D_0$ and  $\cD_r$ given by
\be
\begin{aligned}
	D_0 &= p^\mu\left(\frac{\del}{\del x^\mu} + \Gamma^\rho_{\mu\nu}p_\rho\frac{\del}{\del p_\nu} \right)\\
	\cD_r &= \psi_r^\mu\frac{\del}{\del x^\mu} + p^\mu\frac{\del}{\del\psi^\mu_r}\ ,
\end{aligned}
\ee
where $D_0$ is the same bosonic vector field as before while the $\cD_r$ are fermionic. These vectors obey 
\be
	\left\{\cD_r,\cD_s\right\} = \delta_{rs} D_0\,,
\ee
which is a version of the $\cN=2$ supersymmetry algebra along the super null geodesic.

Similarly to the bosonic case of section~\ref{sec:brief-rev}, we define non--projective super ambitwistor space $\A_S$ to be the quotient of $T^*_{SN}M$ by the action generated by these vectors; it is also the symplectic quotient of $T^*_SM$ by the same action.
\be
	\A_S := T^*_{SN}M / \left\{D_0, \cD_r\right\} \cong T^*_SM\, /\!\!/ \left\{D_0,\cD_r\right\}\,.
\ee
To obtain projective super ambitwistor space $\PA_S$ we further quotient by the Euler vector field so that $\PA_S = \A_S / \{\Upsilon\}$ where
\be
	\Upsilon= 2p_\mu \frac\p{\p p_\mu} + \sum_{r=1}^2 \psi_r^\mu \frac\p{\p\psi^\mu_r}
\ee
is extended to scale the fermionic directions at half the rate it scales the null momentum $p$.  We denote the line bundle $\A_S\rightarrow P\A_S$ by $\cO(-1)$ so that the $\psi_r^\mu$ take values in $\cO(1)$ and $p_\mu$ and the symplectic potential $\theta_S$ take values in $\cO(2)$. We thus identify $\cO(2)$ as the contact line bundle here. Corresponding to~\eqref{doublefibration} in the bosonic case, we now have the double fibrations
\be\lab{susydoublefibration}
\begin{aligned}
\begin{picture}(50,40)
\put(0.0,0.0){\makebox(0,0)[c]{$\A_S$}}
\put(60.0,0.0){\makebox(0,0)[c]{$M$}}
\put(34.0,33.0){\makebox(0,0)[c]{$T^*_{SN}M$}}
\put(5.0,18.0){\makebox(0,0)[c]{$\pi_1$}}
\put(55.0,18.0){\makebox(0,0)[c]{$\pi_2$}}
\put(25.0,25.0){\vector(-1,-1){18}}
\put(37.0,25.0){\vector(1,-1){18}}
\end{picture}\hspace{3cm}
\begin{picture}(50,40)
\put(0.0,0.0){\makebox(0,0)[c]{$\PA_S$}}
\put(60.0,0.0){\makebox(0,0)[c]{$M$}}
\put(34.0,33.0){\makebox(0,0)[c]{$PT^*_{SN}M$}}
\put(5.0,18.0){\makebox(0,0)[c]{$\pi_1$}}
\put(55.0,18.0){\makebox(0,0)[c]{$\pi_2$}}
\put(25.0,25.0){\vector(-1,-1){18}}
\put(37.0,25.0){\vector(1,-1){18}}
\end{picture}
\end{aligned}
\ee
Super ambitwistor space has some additional structure that we will use.  Firstly we have the two involutions $\tau_r$  with $\tau_r \psi_s=(-1)^{\delta_{rs}}\psi_s$, leaving 
$(x^\mu,p_\mu)$ invariant.  These involutions are the $\Z_2\times\Z_2$ symmetry used in the worldsheet action above.   We also have that $g_{\mu\nu}\psi_1^\mu\psi_2^\nu$ descends to $\PA_S$ as a section of $\cO(2)$ generating the $R$-symmetry, although it will not generally be preserved under deformations.

\medskip

As before there is a Penrose transform  between cohomology of $\PA_S$ and fields on space--time. For gravity we will just be concerned with 
$[\delta\theta_S] \in H^{0,1}(\PA_S,\cO(2))$, again thought of as a perturbation of the contact 1-form $\theta_S$.  The principal is much the same as before, but there are some new features we briefly point out here. (A more complete treatment of the Penrose transform in this supersymmetric context may be found in appendix~\ref{P-Transform-RNS}.) Again, to Penrose transform $\delta\theta_S$ to obtain fields on space--time, we first pull it back to give $\pi_1^*(\delta\theta)$ on $PT^*_{SN}M$.  Here it becomes cohomologically trivial as there are no $H^1$s, so we can write $\pi_1^*(\delta\theta)=\delbar j$ where $j$ is determined only up to the addition of a polynomial of weight two in the $\psi_r$ and one in $p_\mu$.  Because it was pulled back from $\PA_S$, all three of the vector fields $D_0$ and $\cD_r$ annihilate $\pi_1^*(\delta\theta)$, so that $D_0j$ and $\cD_rj$ are global and holomorphic and can therefore be expanded as polynomials of the appropriate degree for their weight in $p_\mu$ and $\psi_r^\mu$.  However, because $D_0j = \cD_1^2j = \cD_2^2j$, it is not necessary to consider $D_0j$ itself. We set
\be
	J_r:=\cD_rj\in \cO(3) \, , 
\ee
and the definitions and commutation relations~\eqref{susy} show that the $J_r$ obey
\be
\cD_1J_1=\cD_2J_2\, , \qquad \cD_2J_1+\cD_1J_2=0\, .
\ee
It is easy to see that these relations are solved if there exists a global $U\in\cO(2)$ such that
\be
	J_1=\cD_2 U\qquad \hbox{and} \qquad J_2=-\cD_1 U\, .
\ee
It is more non--trivial to see that there is a choice of the gauge freedom in $j$ so that such a $U$ always exists whenever $\delta\theta$ is invariant under the involutions $\tau_r$.  Imposing also oddness under the $\tau_r$, we must have that 
\be
	H_S=H^{\mu\nu}(x)\psi_{1\mu}\psi_{2\nu}\, 
\ee
for some tensor $H^{\mu\nu}(x)$ that depends only on $x$ but is otherwise arbitrary. In particular, we do not require $H^{\mu\nu}$ to be either symmetric or trace--free.  As in the bosonic case, $H^{\mu\nu}$ also obeys no field equations at this stage. The remaining gauge freedom in $j$ induces the change $\delta H^{\mu\nu}=\del^{(\mu}v^{\nu)}$ for some vector field $v^\mu$ on $M$, so that $H^{\mu\nu}$ is defined modulo diffeomorphisms. 

To describe momentum eigenstates we take $H^{\mu\nu}(x)=\epsilon_1^\mu\epsilon_2^\nu\e^{\im k\cdot x}$ as before, where we have now written the polarization tensor in terms of two vectors $\epsilon_{1,2}^\mu$ as usual. The  corresponding $H_S$ is now given by
\be\lab{HS}
	H_S=\epsilon_1\cdot\psi_1 \ \epsilon_2 \cdot \psi_2\,\e^{\im k\cdot x} 
\ee
whereupon 
\be
	J_1= \epsilon_1\cdot\psi_1\left(\epsilon_2\cdot p +k\cdot \psi_2\ \epsilon_2\cdot\psi_2\right)\e^{\im k\cdot x}
\ee
and $J_2$ is obtained by exchanging $1\leftrightarrow 2$ and including a minus sign. These give 
\be
	j= \frac{\e^{\im k\cdot x}}{k\cdot p}\,\prod_{r=1}^2\left(\epsilon_r\cdot p +k\cdot\psi_r\ \epsilon_r\cdot \psi_r\right)
\ee
and
\be\lab{etasuper}
	\delta\theta = \bar\delta (k\cdot p)\,\e^{\im k\cdot x}\,\prod_{r=1}^2\left(\epsilon_r\cdot p +k\cdot\psi_r\,\epsilon_r\cdot \psi_r\right) \, .
\ee
as the deformation of the super contact structure. It is easy to see that $D_0(\pi_1^*\delta\theta)=\cD_r(\pi^*_1\delta\theta)=0$.   As before, the Penrose transform implies no 
field equations classically, although we will again see that they arise quantum mechanically in the next section.  
Note incidentally that we have a potential $\xi$ for $\delta\theta$ given by
\be
  	\xi =\e^{\im k\cdot x} \epsilon_1\cdot \psi_1\, \epsilon_2 \cdot \psi_2 \bar \delta(k\cdot p)
\ee
which obeys $\cD_1\cD_2\xi = \delta\theta_S$. However, although $\xi$ satisfies $D_0 \xi=0$, it does not satisfy $\cD_r\xi=0$, and so lives on the larger space 
$PT^*_{SN}M/\{D_0\}$ rather than on super ambitwistor space.


\subsection{Quantization}
As before, in order to quantize we introduce the bosonic ghosts $\gamma_r\in \sqrt {T_\Sigma}$, $\beta_r\in K^{3/2}$ as well as the fermionic ghosts $c$ and $\tilde c$ that we had in the bosonic model. In the gauge where $e=\chi_r=0$,  the ghost action takes the standard form
\be\label{ghosts-f}  
	S_{\rm gh}= \frac{1}{2\pi}\int_\Sigma b\delbar c+ \tilde b\delbar \tilde c+\sum_{r=1,2} \beta_r\delbar \gamma_r\ .
\ee
In particular, all sets of ghosts are holomorphic in this chiral model. The BRST operator is extended to become
\be
	Q=\oint cT + \frac{\tilde c}{2}P^2 + \sum_{r=1}^2 \gamma_r P_\mu \Psi_r^\mu +\frac{\tilde b}{2}\gamma_r\gamma_r \, ,
\ee
where $T$ is now the full stress-energy tensor including contributions from both fermions and ghost systems.  This operator generates the gauge transformations
\be\label{BRST-trans}
\begin{aligned}
	\delta X^\mu&=c\p X^\mu+ \tilde c P^\mu+\sum_r\gamma_r \Psi_r^\mu\\
	\delta \Psi_r^\mu&=c\p \Psi_r^\mu + \frac{1}{2}(\del c)\Psi^\mu_r+ \gamma_r P_\mu\\
	\delta P_\mu&= \del(c P_\mu)
\end{aligned}
\ee
reproducing the worldsheet supersymmetries~\eqref{susy} together with worldsheet diffeomorphism invariance. Thus, in the notation of the previous section, the action of $Q$ reduces the worldsheet path integral from being over the space of maps into $PT^*_{SN}M$ down to the space of maps into $\PA_S$. When (the complexification of) $M$ has $d$ (complex) dimensions, we have central charge 
\be
	{\rm c}=2d + \frac d2 +\frac d2 -26+ 11-26 + 11= 3(d-10)
\ee
so, as in the usual RNS string, the critical dimension is ten,  ensuring that $Q^2=0$ at the quantum level.


\subsection{Vertex Operators}
\label{ssec:vertex}

We now construct vertex operators corresponding to gravitational states on space--time. We will content ourselves with discussing the NS sector (for both sets of fermions $\Psi_r$); the Ramond sector is discussed very briefly in section~\ref{Ramond}. 

As before, the fixed vertex operators take the largely standard form
\be
	U = c\tilde c\delta(\gamma_1)\delta(\gamma_2)\,\epsilon_1\cdot\Psi_1\ \epsilon_2\cdot\Psi_2\ ,
\ee
where the ghost insertions $c\tilde c\delta(\gamma_1)\delta(\gamma_2)$ restrict us to considering worldsheet diffeomorphisms and gauge transformations~\eqref{BRST-trans} that act trivially at the insertion point of U, and where the rest of the vertex operator is the field $H_S$ obtained in~\eqref{HS}. These vertex operators thus fix the residual symmetry in the $\cD_r$ directions, enforcing that $\xi = \cD_1^{-1}\cD_2^{-1} j$ is constant at its insertion point. (See section~\ref{sec:superambi} for the definition of $j$ and $\xi$.) They are very similar in appearance to the usual graviton vertex operator of the RNS string, except that here all the fields are holomorphic. In particular, the conformal weight of $c$ and $\tilde c$ is compensated for by the rest of the vertex operator, which transforms as a quadratic differential.

The usual descent procedure in the supersymmetric directions transforms $U$ into the vertex operator
\be
	c\tilde c V = c\tilde c \,\e^{\im k\cdot X}\prod_{r=1}^2\left(\epsilon_r\cdot P +k\cdot\Psi_r\ \epsilon_r\cdot \Psi_r\right)\,.
\ee
As in the bosonic case, this fixed vertex operator enforces the gauge condition $j=$constant at its insertion point, fixing the residual symmetry along $D_0$. Finally, the integrated vertex operator is
\be
	\int_\Sigma\cV = \int_\Sigma  \bar\delta (k\cdot P)\,\e^{\im k\cdot X}\,\prod_{r=1}^2\left(\epsilon_r \cdot P +k\cdot \Psi_r\ \epsilon_r\cdot\Psi_r\right)
\ee
and represents a deformation of the action corresponding to the deformation $\theta_S\to\theta_S+\delta\theta_S$ of the contact structure on $\PA_S$. This is just the supersymmetric version of the contact structure deformation corresponding to a momentum eigenstate on $\PA_S$ as given in~\eqref{etasuper}.

The spectrum arising from these vertex operators includes a graviton in the form of $g_{\mu\nu}=\epsilon_{1(\mu}\epsilon_{2\nu)}\e^{\im k\cdot X}$, together with a scalar dilaton
$\phi=\epsilon_1^\mu\epsilon_{2\mu}\e^{\im k\cdot X}$  and a 2-form $B_{\mu\nu}=\epsilon_{1[\mu}\epsilon_{2\nu]}\e^{\im k\cdot X}$ which we identify as the ten dimensional Neveu-Schwarz $B$-field. Altogether, these fields constitute the NS-NS sector of ten dimensional supergravity. Although classically the vertex operators can be defined off--shell,  in checking BRST closure one meets double contractions whose vanishes enforces the on--shell conditions $k^2=\epsilon_r\cdot k=0$. These conditions are also ensure that the vertex operators themselves are free from normal ordering ambiguities. As before, the $XX$ OPE is trivial so the ambitwistor string spectrum contains are no massive states.

\subsection{Gravitational scattering amplitudes}

At genus zero, $h^1(\Sigma,T_\Sigma) = h^1(\Sigma,T^{1/2}_\Sigma)=0$, so the gauge fields $e$ and  $\chi_r$ may all be set to zero using the gauge transformations~\eqref{extended}-\eqref{susy}. There are three zero modes of each of $c$ and $\tilde c$ as before, and in addition each of the $\gamma_r$ ghosts has two zero modes as in the RNS string. To fix these zero modes we insert two $U$ operators and one $c\tilde cV$ operator so that $n$-particle tree--level amplitudes are given by the correlation function
\be\label{corr}
	\cM(1,\ldots,n)=\left\la U_1U_2\,c_3\tilde c_3 V_3\, \int \cV_4\cdots \int \cV_n \right\ra\,.
\ee
Much of the evaluation of the path integral proceeds as before.  In particular, the $(X,P)$-system may be treated as before and we again find an overall factor of momentum conservation (now in ten dimensions) and that $P_\mu$ is frozen to be
\be
	P_\mu(\sigma) = \rd\sigma\sum_{i=1}^n \frac{k_{i\mu}}{\sigma-\sigma_i}\, .
\ee
Furthermore, the $n-3$ factors of $\bar\delta(k_i\cdot P(\sigma_i))$ combine with the $\tilde c$ ghost zero modes to produce the permutation invariant factor
\be
	\prod_i' \bar\delta(k_i\cdot P(\sigma_i))  = \frac{\sigma_{12}\sigma_{23}\sigma_{31}}{\rd^n\sigma}\,
	\prod_{i=4}^n\,\bar\delta\left(\sum_{j\neq i }\frac{k_i\cdot k_j}{\sigma_i-\sigma_j}\right)
\ee
imposing the scattering equations as before.

The main new ingredient is the contribution from the fermions $\Psi_r$. Each set $r=1,2$ are decoupled both in the action and the vertex operators, so it suffices to treat the contribution from, say, $\Psi_1$. To evaluate the correlator, first consider the path integral
\be\lab{fermioncorr}
	\int[\rd\Psi] \,\exp\left(-\frac{1}{2\pi}\int_\Sigma \Psi_\mu\delbar\Psi^\mu\right)\,\prod_{i=1}^n \epsilon_i\cdot\Psi(\sigma_i)\,k_i\cdot\Psi(\sigma_i)\ .
\ee 
It is a standard result that~\eqref{fermioncorr} yields the Pfaffian of the $2n\times 2n$ antisymmetric matrix 
\begin{equation}
	M'= \begin{pmatrix}
		\,A & -C'^{\rm T}\\
		\,C' & B
		\end{pmatrix}\ ,
\label{Mdef1}
\end{equation}
where the $n\times n$ matrices $A$, $B$ and $C'$ have entries
\begin{equation}
	A_{ij} = k_i\cdot k_j\frac{\sqrt{\rd\sigma_i\rd\sigma_j}}{\sigma_{ij}}\qquad B_{ij} = \epsilon_i\cdot\epsilon_j\frac{\sqrt{\rd\sigma_i\rd\sigma_j}}{\sigma_{ij}} 
	\qquad C'_{ij} = \epsilon_i\cdot k_j\frac{\sqrt{\rd\sigma_i\rd\sigma_j}}{\sigma_{ij}}
\label{Mdef2}
\end{equation}
for $i\neq j$,  and 
\be
 A_{ii}= B_{ii}= C'_{ii}=0\,.
\ee  
These entries result from contracting either the $\epsilon\cdot\Psi$ or the $k\cdot\Psi$ at site $i$ to the $\epsilon\cdot\Psi$ or $k\cdot\Psi$ at site $j$. As usual, we get the Pfaffian of $M'$ rather than its determinant because the action in~\eqref{fermioncorr} is quadratic in $\Psi$, rather than bilinear in $\Psi$ and $\bar\Psi$.

Now, the form of the vertex operators means we must actually consider a product of terms of the form $(\epsilon_i\cdot P(\sigma_i)+ \epsilon_i\cdot \Psi(\sigma_i) k_i\cdot \Psi(\sigma_i))$.     The additional $\epsilon_i\cdot P(\sigma_i)$ can be incorporated by notionally replacing the vanishing contraction between $ \epsilon_i \cdot \Psi(\sigma_i)$ and $ k_i\cdot \Psi(\sigma_i)$ with 
\be
	\epsilon_i\cdot P(\sigma_i)=\rd\sigma_i\sum_{j\neq i} \frac{\epsilon_i\cdot k_j}{\sigma_{ij}} \, ,
\ee
where we have used the fact that $P(\sigma)$ is frozen by the $X$ path integral. These factors are incorporated into the Pfaffian by replacing the matrix $C'$ by a matrix $C$ whose off--diagonal entries agree with those of $C'$, but where now
\be
	C_{ii}=\epsilon_i\cdot P(\sigma_i)=-\rd\sigma_i\sum_{j\neq i} \frac{\epsilon_i\cdot k_j}{\sigma_{ij}} \, .
\ee

In fact, worldsheet supersymmetry~\eqref{susy} means that the Pfaffian of the $n\times n$ matrix $M = \begin{pmatrix} A & -C^{\rm T}\\ C & B\end{pmatrix}$ vanishes to second order. Our actual correlation function~\eqref{corr} does not have $n$ integrated vertex operators, but rather involves two vertex operators $U$ at sites 1 and 2. These $U$ operators do not contain a factor of $k\cdot\Psi$. In this case, the path integral over $\Psi$ instead leads to ${\rm Pfaff}({M}^{12}_{12})$,  the Pfaffian of the matrix ${M}^{12}_{12}$ obtained by removing the first two rows and columns from $M$. The $U$ operators also involve a $\delta$-function $\delta(\gamma)$ in the ghosts that are responsible for fixing the residual worldsheet supersymmetry by forcing the supersymmetry variations to vanish at these insertion points. Upon performing the $\beta\gamma$ path integral, these $\delta$-functions produce a factor of $\sqrt{\rd\sigma_1\rd\sigma_2}/\sigma_{12}$ coming from the two elements of $H^0(\Sigma,T_\Sigma^{1/2})$. Thus, overall the fermions yield a contribution to the path integral of
\be
	{\rm Pf}'(M):=\frac{\sqrt{\rd\sigma_1\rd\sigma_2}}{\sigma_{12}}\, {\rm Pfaff}({M}^{12}_{12})
\ee
which transforms as a section of $K$ at each of the $n$ marked points. It was shown in~\cite{Cachazo:2013hca} that this factor is indeed permutation invariant -- of course, from the current perspective this is just a consequence of having the freedom to fix the residual worldsheet supersymmetry in any way we choose. Note in particular that since both $U$ and $V$ each do involve a factor of $\epsilon\cdot \Psi$, this operator appears at every site and so no matter where we place the $U$ operators we never remove any rows and columns from $B$, as was necessary in~\cite{Cachazo:2013hca}.

Combining all the pieces, including both sets of fermions $\Psi_r$ and their associated ghosts, we obtain finally the amplitude
\be
	\cM(1,\ldots,n)=\delta^{10}\left(\sum k_i\right) \int \frac{1}{\rm Vol\,SL(2;\C)} \mathrm{Pf}'(M_1)\mathrm{Pf}'(M_2) {\prod_i}'\,\bar\delta(k_i\cdot P(\sigma_i))\, ,
\ee
where $M_1$ is built out of the polarization vectors $\epsilon_{1i}$ and $M_2$ out of the $\epsilon_{2i}$ and where $P(\sigma) = \rd\sigma \sum_i k_i / (\sigma-\sigma_i)$. The two ${\rm Pf}'$s together provide a quadratic differential at each marked point, which becomes a (1,0)-form upon multiplication by $\prod' \bar\delta(k_i\cdot P(\sigma_i))$. Dividing by Vol SL$(2;\C)$ then transforms this to a holomorphic $n-3$ form which may be integrated over a middle dimensional cycle in the moduli space $\cM_{0,n}$ of marked rational curves. This is exactly the expression originally discovered in~\cite{Cachazo:2013hca} and describes all tree--level scattering amplitudes of massless states in the NS--NS sector of pure (type II) supegravity in ten dimensions. 


\section{Yang-Mills amplitudes} 
\label{sec:heterotic}

To construct amplitudes for Yang-Mills fields from ambitwistor strings, we will replace one set of $\Psi$ fields by a more general level k current algebra.  This is somewhat analogous to a heterotic string, although we stress again that all our worldsheet fields will be holomorphic (or left--moving). Thus we have the same fields as before but now with just $r=1$, together with a current  $J_a(\sigma)\in K_\Sigma\otimes \frak{g}$ with OPE
\be\lab{currOPE}
	J_a(\sigma) J_b(\sigma')=\frac{{\rm k}\,\delta_{ab}}{(\sigma-\sigma')^2} + \frac{f_{ab}^cJ_c}{\sigma-\sigma'}+ \cdots\ ,
\ee
where $f^c_{ab}$ are the structure constants for the gauge group $G$, with $a$ a Lie algebra index. As usual, the current algebra could be realized in many ways, such as a free fermionic model or a WZW model. We will not need to be specific. 

The matter action is
\be
	S_{\rm het} =  S_{\rm current}+\frac{1}{2\pi}\int_\Sigma\,P_\mu\delbar X^\mu + \frac{1}{2}\Psi_\mu\delbar\Psi^\mu +\frac{e}{2} P^2+ \chi  P_\mu\Psi^\mu  
\label{hetact}
\ee
where $S_{\rm current}$ is the action for the current algebra and the other fields have the same meaning as before. This model has only one copy of the worldsheet supersymmetry and the BRST operator becomes
\be
Q_{\rm het}=\oint  c T+ \frac{\tilde c}{2}P^2+\gamma P\cdot\Psi +\frac{\tilde b}2 \gamma^2\ ,
\label{Qhet}
\ee
where we now have only one set of $\beta\gamma$ ghosts, and where the holomorphic stress tensor $T$ includes a contribution from the current algebra. This BRST operator implements the symplectic quotient of $T^*M$ generated by $\cD=\{D_0,D_1\}$. 

Unlike the usual heterotic string, because all the fields are chiral, it is possible to balance the central charge of the current algebra against that of the rest of the matter and ghosts and obtain cancellation even away from ten dimensions. The total central charge vanishes provided only the central charge c of the current algebra and the (complex) dimension $d$ of the target space are related as
\be
	{\rm c}=41-\frac{5}{2}d\ .
\label{hetc}
\ee
For example, this gives the standard result ${\rm c}=16$ in ten dimensions, but also allows ${\rm c}=31$ when $d=4$. The possibility of constructing this theory in various dimensions is striking. We note again that both ambitwistor space and the tree--level formulae of~\cite{Cachazo:2013gna,Cachazo:2013hca,Cachazo:2013iea} make sense in any number of dimensions. Of course, modular invariance may be expected to impose strong restrictions on the admissible current algebras at higher genus; we will return to consider these constraints in a subsequent paper. 


\subsection{Yang-Mills amplitudes}

An (off--shell) Yang-Mills bundle on space--time is equivalent to a holomorphic vector bundle $E\to\PA$ on ambitwistor space. To describe perturbative gluons, we consider deformations of the complex structure of this bundle, represented by $\cA_a\in H^1(\PA,{\rm End}(E))$.
Essentially by definition, the deformation of the worldsheet current algebra action is
\be\label{def-current}
 	\int_\Sigma\cV^a=\int_\Sigma  \cA_aJ_a \ ,
\ee
which may be interpreted as the integrated vertex operator for a gluon with ambitwistor wavefunction $\cA_a$. To describe a momentum eigenstate with polarization vector $\epsilon_\mu$, we choose the wavefunctions 
\be
	\cA^a =\bar\delta(k\cdot P)\,\e^{\im k\cdot X} (\epsilon\cdot P+\epsilon\cdot \Psi k\cdot\Psi)  \,T^a
\ee
as in~\eqref{susy-dolrep}, where $T^a\in \frak{g}$ labels the colour of the external state. The integrated vertex operator thus becomes 
\be\lab{int-YM-V-op}
	\cV^1  = \bar\delta(k\cdot P) \left[\epsilon\cdot P + \epsilon\cdot\Psi\,k\cdot \Psi\right]\,\e^{\im k\cdot X}\, T^aJ_a
\ee
and transforms as a (1,1)-form on $\Sigma$. The fixed vertex operators for gluons are 
\be\lab{fix-YM-V-op}
	U^1= c\tilde c\,\delta(\gamma)\, \epsilon\cdot \Psi\,\e^{\im k\cdot X}\, T^aJ_a
\ee
and are worldsheet scalars as expected. The form of the Yang-Mills vertex operators are thus very closely related to the Yang-Mills vertex operators in the standard heterotic string, with differences arising as in the bosonic and type II ambitwistor strings because all the fields are chiral. As usual, these vertex operators are BRST invariant classically for any $k$ and $\epsilon$, but quantum corrections mean BRST closure fails unless $k^2=0$ and $\epsilon\cdot k=0$. If $\epsilon_\mu\propto k_\mu$ then~\eqref{int-YM-V-op} and~\eqref{fix-YM-V-op} are BRST exact. Thus nontrivial vertex operators correspond to on--shell gluons.

To compute the scattering of these Yang-Mills states we again need two $U$ insertions to fix the two $\gamma$ zero modes, one $c\tilde c V$ to fix the remaining $c$ and $\tilde c$ zero-mode and then the rest of the vertex operator insertions must be  $\cV$s.  Thus we consider
\be\lab{Mhet}
	\cM_{\rm het} (1,\ldots ,n)=\left\langle U^1_1\, U^1_2\, c_3\tilde c_3 V^1_3\int \cV^1_4 \cdots \int \cV^1_n\right\rangle \, .
\ee
The current algebra is decoupled from the $\Psi$ and $XP$ system in $S_{\rm het}$, much of the calculation proceeds as in the type II case. In particular, the path integral over the $\Psi$ field and ghosts gives the Pfaffian as before, though now only one copy.  In all, the path integral~\eqref{Mhet} may be evaluated as
\be
	\delta^d\left(\sum_i k_i\right)\int \frac{\rd^n\sigma}{\rm Vol\,SL(2;\C)}{\prod_{i}}'\,\bar\delta(k_i\cdot P(\sigma_i)) \ {\rm Pf}'(M)\
	\left[\frac{\tr(T_1T_2\cdots T_n)}{\sigma_{12}\sigma_{23}\cdots \sigma_{n1}} \ +\  \cdots\ \right]\ ,
\label{hetamp}
\ee
where the term in square brackets arises from the current correlator. Here, the ellipsis represents a sum over both non-cyclic permutations of the marked points and also multi-trace contributions. The ${\rm Pf}'$ and the current algebra provide a quadratic differential at each marked point, which combines with the $\delta$-functions imposing the scattering equations and the $1/{\rm Vol\,SL(2;\C)}$ factor to produce a holomorphic $n-3$ form that may be integrated over a middle dimensional slice of $\cM_{0,n}$. The leading trace terms in~\eqref{hetamp} coincide exactly with the representation of all Yang-Mills tree amplitudes found in~\cite{Cachazo:2013hca}.

The multi--trace terms  are indicative of coupling to gravity, with the gravitational contribution linking the gauge singlets as in the standard heterotic string.   Indeed, this model also contains the (fixed) vertex operator
\be\lab{fix-gr-V-op-het}
	c\tilde c \delta(\gamma)\left(H_{\mu\nu}P^\mu\Psi^\nu + C_{\mu\nu\rho}\Psi^\mu\Psi^\nu\Psi^\rho\right)\,\e^{ik\cdot X}
\ee
(and its associated integrated operator) that describes gravitational states (metric + B-field + dilaton) with polarization $H_{\mu\nu}$, together with a 3-form potential $C$.  Again, in order for these to be BRST invariant vertex operators quantum mechanically, we need $k^2=k^\mu C_{\mu\nu\rho}=k^\mu H_{\mu\nu} = k^\nu H_{\mu\nu}=0$. However, because of the presence of the 3-form field $C$, we no longer have the appropriate spectrum for the NS sector of heterotic gravity. Furthermore, as in the bosonic case, the amplitudes obtained by scattering these states do not agree with those of gravity, even if we turn off $C$.


\subsection{Scalar fields from an additional current algebra}

In~\cite{Cachazo:2013iea} the authors constructed amplitudes for massless scalars transforming in the adjoint of some gauge group $G\times \widetilde G$. We can duplicate these here if we introduce a further level $\tilde {\rm k}$ set of currents $\tilde J_{\tilde a}\in K_\Sigma\otimes\tilde {\frak g}$ in place of the remaining $\Psi$ fields. There is thus no remaining worldsheet supersymmetry and the BRST operator is the same as the bosonic case~\eqref{Q-bos}, but with the stress tensor including those of the current algebras. Each of the $J_a$ and $\tilde J_{\tilde a}$ currents have the standard OPE~\eqref{currOPE}, while $J_a(\sigma)\tilde J_{\tilde a}(\sigma')\sim 0$. The central charge vanishes provided the contributions c and $\tilde{\rm c}$ from the current algebras obey
\be
	{\rm c}+\tilde{\rm c} = 2(26-d) 
\ee
for a $d$ complex dimensional space--time.

In order to construct amplitudes, we introduce the (1,1)-form vertex operator $\cV^0=\bar\delta(k\cdot P)J_aT^a\tilde J_{\tilde a}\tilde T^{\tilde a}\e^{\im k\cdot X}$. Integrating this vertex operator over the worldsheet provides a deformation to the action that now couples the two current algebras. Via the ambitwistor Penrose transform, the contribution $\bar \delta (k\cdot P)\e^{ik\cdot X}T^a\tilde T^{\tilde a}$ to the integrated vertex operator is an ambitwistor space representative of the scalar field $\phi^{a\tilde a}=\e^{ik\cdot X}T^a\tilde T^{\tilde a}$ on space--time (see appendix~\ref{ambi-app}).  We also have the fixed vertex operator $c\tilde c V^0 = c\tilde c J_aT^a\tilde J_{\tilde a}\tilde T^{\tilde a}\e^{\im k\cdot X}$ obtained by the Penrose transform as before (see section~\ref{sec:brief-rev} and appendix~\ref{ambi-app}). For these operators to be $Q$-invariant, we require that $k^2=0$. 

Since the two current algebras commute, their path integrals may be performed independently of eachother (an independently of the $XP$ system). Each factor leads to both single trace and multi-trace terms.  Picking out only the leading trace contributions from each factor, we find
\be
\begin{aligned}
	&M_{\rm scal}(1,\ldots ,n)=\left\langle c_1\tilde c_1V_1^0\, c_2\tilde c_2 V_2^0\, c_3\tilde c_3 V_3^0\, \int \cV_4^0\cdots \int \cV_n^0\right\rangle \\
&\ =\delta^d\!\left(\sum_ik_i\right)\int \frac{(\rd^n\sigma)^2}{\mathrm{Vol\,SL}(2,\C)} 
{\prod_i}'\, \bar\delta(k_i\cdot P(\sigma_i))\left[\frac{\tr (T_1\cdots T_n)}{\sigma_{12}\sigma_{23}\cdots \sigma_{n1}} 
\times \frac {\tr(\tilde T_{\alpha(1)}\cdots \tilde T_{\alpha(n)})}{\sigma_{\alpha(1)\alpha(2)}\cdots \sigma_{\alpha(n)\alpha(1)}}+\cdots \,\right]
\end{aligned}
\ee
where the ellipsis denotes both non--cyclic permutations of this `double' leading trace term, together with multi--trace terms.  Again, the quadratic differentials from the two holomorphic current algebras combine with the $\delta$-functions imposing the scattering equations and the 1/Vol SL$(2;\C)$ to produce a holomorphic $n-3$ form that may be integrated over a real slice of $\cM_{0,n}$.  The double leading trace part coincides with the scalar field scattering formulae of~\cite{Cachazo:2013iea}. The sum over permutations of this double leading trace term is there argued to give the tree--level amplitudes corresponding to the space--time scalar field theory with action
\be
	S[\phi^{a\tilde a}]=\int_M \frac{1}{2}\del_\mu\phi^{a\tilde a}\del^\mu\phi_{a\tilde a}
	+ \frac{1}{3}f_{abc}\tilde f_{\tilde a\tilde b\tilde c}\phi^{a\tilde a}\phi^{b\tilde b}\phi^{c\tilde c}\, .
\ee
However,  the string theory also generates multi--trace contributions in the correlator. These perhaps arise from coupling the scalar to gravity in this bosonic string, but are not straightforward to interpret.


\section{Conclusions and further directions}
\label{sec:conclusions}

We have presented worldsheet models whose $n$-point correlation functions at genus zero reproduce the new representations of tree--level gravitational, Yang-Mills and scalar 
amplitudes presented in~\cite{Cachazo:2013hca}. These representations
are supported on solutions of the scattering equations~\eqref{scatt}
by virtue of the origin of the wave functions as cohomology classes on
ambitwistor space. The amplitudes for particles of different spin came from different string theories, with the scalar, Yang-Mills and gravitational amplitudes arising from the bosonic, `heterotic' and `type II' ambitwistor strings, respectively. The bosonic and heterotic models are problematic because the gravitational amplitudes they contain do not seem to correspond to Einstein gravity. (Indeed, we are not yet certain whether their amplitudes agree with any known space--time theory of gravity.) However, the type II model does seem to be consistent.

As noted in~\cite{Cachazo:2013iea,Cachazo:2013hca}, one of the most intriguing features of these scattering equations is that they also determine saddle points in the usual 
string worldsheet path integral which dominate the limit of high energy, fixed angle scattering studied by Gross \& Mende~\cite{Gross:1987ar}. Classical gravitational and 
Yang-Mills amplitudes emerge from string theory when the energy scales are small compared to the string tension, while the Gross--Mende limit is the opposite case where all 
kinematic invariants are very large. It is remarkable that the same equations determine both limits. We hope that the present derivation of the amplitude representations of~\cite{Cachazo:2013iea} from a worldsheet model not too distant from the usual RNS string helps provide a starting point to understand this fascinating connection.  

\bigskip

We conclude this final section by listing a few possible avenues that seem ripe for further investigation.


\subsection{Ramond sector vertex operators}
\label{Ramond}

The type II ambitwistor string appears to be equivalent to a type II supergravity in 10 dimensions.  To be sure of this we need to see that, as well as the NS-NS\footnote{In our purely chiral context, by the NS-NS sector, we mean the Neveu-Schwarz sector for each of the two sets of left moving worldsheet fermions $\Psi_r$.} sector studied in this paper, it also correctly reproduces the (massless) Ramond-Ramond and Ramond-NS sectors.  The formulation of these ambitwistor strings is sufficiently close to the standard RNS string that we expect standard technology can be brought to bear.  

In particular, we anticipate that the model also contains two space--time gravitinos, associated to the vertex operator
\be
	\int_\Sigma \bar\delta(k\cdot P)\, \cV_1^\alpha\,\delta(\gamma_2)\epsilon^\mu_{\alpha}\left(P_\mu + \Psi_{2\mu} \ k\cdot\Psi_2\right)\e^{\im k\cdot X}
\ee
and a similar one obtained by exchanging $\Psi_1 \leftrightarrow\Psi_2$. Here, $\cV_1^\alpha = \e^{\phi/2} P_\mu\gamma^\mu_{\alpha\beta}  \Theta_1^\beta \in K_\Sigma$, where $\phi$ arise in the bosonization of the $\beta\gamma$ ghost system, $\gamma_{\alpha\beta}^\mu=\gamma^\mu_{(\alpha\beta)}$ are the ten dimensional Van der Waerden symbols, and $\Theta_1^\alpha$ is the spin field for the $\Psi_1$ system (see {\it e.g.}~\cite{Friedan:1985ge,Polchinski:1998rr}). There are likewise Ramond-Ramond sector $p$-form fields created by vertex operators
\be
	\int_\Sigma  \bar\delta(k\cdot P) \cV_1^\alpha \cV_2^\beta\gamma_{\alpha\beta}^{\mu_1\ldots\mu_p}\epsilon _{\mu_1\ldots\mu_p}\e^{\im k\cdot X}
\ee
that involve spin fields for both the $\Psi_r$ systems. Once more, the presence of the $\bar\delta(k\cdot P)$ term is dictated by the Penrose transform, and is necessary to construct well--defined vertex operators in the case where both sets of worldsheet fermions are holomorphic. It will be fascinating to see whether the amplitudes involving these fields indeed agree with those of supergravity, and what constraints on these vertex operators are imposed by modular invariance.


\subsection{Loop amplitudes}
\label{sec:1loop}

One advantage of understanding the expressions found in~\cite{Cachazo:2013iea} from the perspective of a worldsheet theory is that it provides a natural way to try to extend  these amplitudes beyond tree--level: we simply consider the relevant correlation function on a higher genus Riemann surface.  One might have said the same also for Witten's original twistor string, and also for the twistor string developed by one of us~\cite{Skinner:2013xp} for $\cN=8$ supergravity.  However, the ambitwistor strings are appreciably closer to the standard RNS string, so it is likely that one can make more rapid progress with the current model. We note however that ten dimensional supergravity is UV divergent even at one loop. It will be interesting to see how this arises from the current models.


\subsection{Green-Schwarz strings in ambitwistor space}
\label{other}

Here we have focussed on the bosonic and RNS string, but our general philosophy applies equally well to models with manifest space--time supersymmetry. One simply views ambitwistor superspace as the space of super null geodesics in superspace, in the original spirit of Witten~\cite{Witten:1985nt}. Alternatively, in four dimensions, one may make space--time supersymmetry manifest using the close relation between ambitwistors and ordinary twistors. We now briefly survey such models, modelling our discussion on that given by Berkovits for standard string theory \cite{Bedoya:2009np}.

The Green-Schwarz models can be motivated by starting from the Brink-Schwarz superparticle \cite{Brink:1981nb} for a null geodesic  in $(10|16)$-dimensional superspace with coordinates $(x^\mu, \theta^\alpha)$.  Its action is
\be
	S=\int  P_\mu (\rd X^\mu  - \gamma^\mu_{\alpha\beta}\theta^\alpha\rd\theta^\beta) - \frac{1}{2}e  P^2\, ,
\ee
where $\gamma^\mu_{\alpha\beta}$ is one of the Van de Waerden symbols that arise from decomposing the gamma matrices into their chiral parts.  
As in the RNS case, this can be elevated to an ambitwistor string action
\be\lab{GSact}
S[X,\theta,P]= \int_\Sigma P_\mu (\delbar X^\mu- \gamma_{\alpha\beta}^\mu\theta^\alpha\delbar \theta^\beta) - \frac{1}{2}e P^2
\ee
for fields $(X,\theta):\Sigma \rightarrow \C^{10|16}$, $P\in K\otimes \C^{10}$ and $e$ a Beltrami differential. Exactly as before, this action is manifestly reparametrization invariant and $e$ is a worldsheet gauge field for transformations $\delta X^\mu = \alpha P^\mu$, $\delta\theta^\alpha=0$, $\delta P_\mu=0$, and $\delta e = \delbar\alpha$, with $\alpha$ a worldsheet vector pointing in the holomorphic directions. 

As usual, \eqref{GSact} is invariant under the space--time supersymmetry transformations $\delta X^\mu = \gamma^\mu_{\alpha\beta}\epsilon^\alpha\theta^\beta$, 
$\delta \theta^\alpha = \epsilon^\alpha$ and $\delta P_\mu=\delta e=0$, with $\epsilon^\alpha$ a constant anticommuting parameter. There is also a local $\kappa$-symmetry that arises because, when $P_\mu$ is null, the matrix $P_\mu\gamma^\mu_{\alpha\beta}$  has an 8-dimensional kernel so that the action is degenerate in the fermionic variables. Specifically,  if $\kappa^\alpha$ satisfies $P_\mu \gamma^\mu_{\alpha\beta}\kappa^\alpha=0$, the action is invariant under $\delta\theta=\kappa$.  Thus~\eqref{GSact} really defines a string theory into Witten's version \cite{Witten:1985nt} of superambitwistor space for 10 dimensional space--time, in which a super null geodesic is the $(1|8)$-dimensional supersymmetric extension of the standard light--ray given parametrically by $(X_0+\tau P,\theta^\alpha_0+ \kappa^\alpha)$ where the parameters $(\tau, \kappa^\alpha)\in \C^{1|8}$ satisfy $P_\mu\gamma^\mu_{\alpha\beta}\kappa^\beta=0$.

The conventional Green-Schwarz action is usually quantized in light--cone gauge, breaking manifest covariance,  whereas computing amplitudes in the RNS string requires breaking manifest space--time supersymmetry and the introduction of rather awkward spin fields to describe space--time fermions. These considerations led Berkovits to introduce the pure spinor superparticle and string.  We expect that our procedure should also be applicable to the pure spinor formulation of the superparticle, leading to a pure spinor variant of the ambitwistor string. 


\subsection{Twistor and ambitwistor strings}\label{sec:twistor-strings}

In four dimensions, ambitwistor space -- the space of complex null geodesics -- is closely related to both standard twistor space and its dual. Indeed, the name `ambitwistor' originates with this relation. In four dimensions, a null momentum $p$ can be written as a simple bispinor $p_{\alpha\dot\alpha} = \lambda_\alpha\tilde\lambda_{\dot\alpha}$, where $\lambda_\alpha$ and $\tilde\lambda_{\dot\alpha}$ are each two component spinors. Given a null geodesic with momentum $\lambda_\alpha \tilde \lambda_{\dot\alpha}$ through the point $x$, we can introduce a twistor $Z\in\C^4$ and a dual twistor $W\in\C^4$ by
\be
	W_a=(\lambda_\alpha,\mu^{\dot\alpha})=(\lambda_\alpha,-\im x^{\alpha\dot\alpha}\lambda_{\alpha})\in \T^*\, , 
	\qquad Z^a=(\tilde \mu^\alpha,\tilde\lambda_{\dot\alpha})=(\im x^{\alpha\dot\alpha},\tilde\lambda_{\dot\alpha})\in \T\, .
\ee
It is easily seen that if $(Z,W)$ arise from a null geodesic in this way, then they satisfy
\be
	Z^a W_a=0\ , \qquad \mbox{where} \qquad Z^a W_a = \lambda_\alpha \tilde \mu^{\alpha}+ \mu^{\dot\alpha}\tilde \lambda_{\dot\alpha}\ .
\ee
Conversely, if $Z\cdot W=0$ then $(Z,W)$ arises from such a null geodesic.  The pair $(Z,W)$ has two scalings, one for $Z$ and one for $W$.  The product scaling is clearly that of the original null geodesic, but $\Upsilon=Z^a\frac{\del}{\del Z^a}-W_a\frac{\del}{\del W_a}$ is redundant.  Thus we arrive at the description of ambitwistor space as a symplectic reduction 
\be
	\A_0=\{(Z,W)\in \T\times \T^*\ |\  Z\cdot W=0\}\ /\ \Upsilon\ ,
\ee
where we start with the holomorphic symplectic form $\omega=\rd W_a \wedge \rd Z^a$ and symplectic potential $\theta=W_a\rd Z^a$.

In four dimensions, ambitwistor superspace can likewise be introduced by starting with super null geodesics in $\C^{4|16}$ with coordinates 
$(x^{\alpha\dot\alpha}, \theta_A^\alpha, \tilde \theta^{A\dot\alpha})$, where $A=1,\ldots,\cN$ is an $R$--symmetry index.  A super null geodesic is the $(1|2\cN)$-dimensional 
subspace described parametrically as $(x_0^{\alpha\dot\alpha}+ \tau \lambda^\alpha\tilde\lambda^{\dot\alpha}, \ \theta_{0A}^{\alpha}+\lambda^\alpha\kappa_A,\ 
\tilde\theta_0^{A\dot\alpha}+\tilde\lambda^{\dot\alpha}\tilde\kappa^A)$ where the $\kappa$ and $\tilde \kappa$ are anticommuting parameters\footnote{For $\cN=4,8$ this can be 
understood by reduction from 10 dimensional $\kappa$-symmetry.}. Given a super null geodesic we can define a supertwistor and dual supertwistor, each in $\C^{4|\cN}$, by
\be
	\cW_I:=(W_a,\chi_A)=(W_a,\theta_{0A}^\alpha\lambda_\alpha)\, , \qquad \cZ^I:=(Z^a,\tilde \chi^A)=(Z^a, \tilde \theta_0^{A\dot\alpha}\tilde \lambda_{\dot\alpha})\ .
\ee
Again, if the supertwistor arises in this way we will have $\cZ\cdot \cW:= Z^a W_a + \tilde\chi^A\chi_A=0$. We can therefore define superambitwistor space by
\be
	\A=\{(\cZ^I,\cW_I)\in \C^{4|\cN}\times\C^{4|\cN}\ |\ \cZ\cdot\cW=0\}\ /\ \Upsilon\ ,
\ee
where $\Upsilon$ is extended to also scale the fermionic directions in the obvious way. Again it is a symplectic quotient, now by $\cZ\cdot \cW$. For $\cN=3$ the projectivisation $P\A$ turns out to be a Calabi-Yau supermanifold.  This space was introduced by Witten~\cite{Witten:1978xx}, who showed that on--shell $\cN=3$ super Yang-Mills fields correspond to deformations of trivial holomorphic vector bundles on this space.  

In terms of these coordinates, an ambitwistor superstring can be obtained by gauging the constraint $\cZ\cdot\cW=0$. We thus have the worldsheet action
\be
	S[\cZ,\cW,A]=\int_\Sigma \cW_I(\delbar +A)\cZ^I \ ,
\ee 
where $\bar\del+A$  defines a $\delbar$-operator a line bundle $\cL\rightarrow \Sigma$ such that
\be
	\cZ:\Sigma\rightarrow \cL\otimes \C^{4|\cN}\, , \qquad \cW:\Sigma \rightarrow \tilde\cL\otimes\C^{4|\cN} 
\ee
where $\cL\otimes\tilde\cL\cong K$. When $\cN=4$, this is essentially a chiral version of  Berkovits' formulation of twistor strings. Similarly, the $\cN=8$ twistor string of \cite{Skinner:2013xp} can be understood as belonging to this general family of ambitwistor strings (albeit with additional fields that we do not discuss here).  However, the symmetrical presentation now allows us to consider line bundles $\cL$ of negative as well as positive degree and vertex operators that depend non--trivially on $\cW$ as well as $\cZ$.  From this point of view, with maximal supersymmetry we have a doubling of the degrees of freedom unless further constraints (perhaps involving a real structure) are imposed.


\bigskip

\noindent  {\bf Acknowledgements:} We thank Tim Adamo, Malcolm Perry and especially Edward Witten for helpful discussions. DS is supported by an IBM Einstein Fellowship of the IAS and by The Ambrose Monell Foundation.  LJM is supported by EPSRC grant EP/J019518/1 and the Simons Center for Geometry for geometry and Physics. The research leading to these results has received funding from the European Research Council under the European Community's Seventh Framework Programme (FP7/2007-2013) / ERC grant agreement no. [247252].

\bibliography{ATSbib}
\bibliographystyle{JHEP}

\appendix

\section{The bosonic Ambitwistor string as  an $\alpha'\rightarrow 0$ limit}\label{sec:alpha0}
As a heuristic motivation, we derive \eqref{boson-str} as a chiral $\alpha'\rightarrow 0$ limit of the standard bosonic string.  We write the Polyakov string action for a map $X:\Sigma\rightarrow M_\R$ from the Riemann surface $\Sigma$ to a real $d$-dimensional space-time $(M^d_\R,g)$ as
\be
S=\frac1{2\pi \alpha'} \int_\Sigma \frac1{\sqrt{1- {e}'\bar {e}'}}(\p X\cdot \delbar  X + {e}' \p X\cdot \p X + \bar  {e}'  \delbar X\cdot \delbar X)\, .
\ee
Here the dot denotes inner product with respect to the metric $g$ on $M$, and the metric $h$ on $\Sigma$ has been referred to a background choice of complex structure $\delbar=d\bar\sigma \p_{\bar\sigma}$ by $h^{ij}\p_i\p_j=\Omega (\p_\sigma\bar \p_{\bar \sigma} + {e}'\p_\sigma\p_\sigma+ \bar {e}\p_{\bar\sigma}\p_{\bar\sigma}$.  We now take $\alpha'\rightarrow 0$  by introducing Lagrange multipliers $P$ and $\tilde P$, and rescalings  ${e}= \alpha' {e}'$, $\tilde  {e}=\alpha^{\prime -2}\bar {e}'$ to obtain the equivalent action
\be
S= \frac1{2\pi}\int_\Sigma \frac1{\sqrt{1-\alpha' {e}\bar {e}}}(P\cdot \delbar X +\alpha' \bar P \cdot \p X  -\alpha^{\prime 2} P\cdot \bar P  +  {e} P^2 + \alpha^{\prime }\bar  {e} \bar P^2)
\ee
as can be seen by eliminating $P$ and $\bar P$ (here $P^2=P\cdot P)$ etc.). Taking $\alpha'\rightarrow 0$ we obtain the bosonic classical string action
\be\label{bos-action}
S_b=\frac1{2\pi}\int_\Sigma P\cdot \delbar X+  {e} P^2\, .
\ee


\section{Ambitwistor space and the Penrose-Ward transform.}
\label{ambi-app}
Here we give a few more technical details on the Penrose transform
between linear fields on space-time and cohomology classes on
Ambitwistor space, both in the bosonic case, and in the case where we
have just one $\Psi$ (the heterotic case).  When we come to the
Penrose-Ward transform we will want to work on its projectivisation,
$P\A$.  Ambitwistor space in general has cohomology in degree 1 and
$d-2$, but here we only discuss degree 1 as that is the
only case needed in this work although conceivably a role for the higher degree
cohomology might emerge at some point.

\subsection{The bosonic case}

The Penrose transform can be described for $H^1$s with values in
$L^n$ for all $n$ as follows.

\begin{thm}
The Penrose transform maps cohmology classes on $P\A{}$ to fields on spacetime as follows.  
For $n\geq -1$ we have 
\be\label{Ptrans}
H^1(\PA,L^n)= \{A_{\mu_0\ldots \mu_n}=\phi_{(\mu_0\ldots \mu_{n})_0}\}/\{\nabla_{(\mu_0} a_{\mu_1\ldots \mu_n)_0}\}\, .
\ee
Here $(\ldots )_0$ denotes `the symmetric trace-free part'. When $n<-1$ $H^1(\PA ,L^n)=0$.
\end{thm}
\noindent
{\bf Proof:}
Homogeneity degree $n$ functions $\cO_{P\A{}}(n)$ on ambitwistor space can be represented as homogeneous degree $n$ functions on the projective cotangent bundle $PT^*M$ restricted to $P^2=0$ that are anniilated by the geodesic flow $D_0$.  Thus we have the short exact sequence:
\be\label{SES}
0\rightarrow L^n_{\PA } \rightarrow L^n_{ PT^*_NM } \stackrel{D_0}{\rightarrow} L^{n+1}_{PT^*_NM}\rightarrow 0\, .  
\ee
The associated long exact sequence in cohomology degenerates quickly because the cohomology of the projective lightcone vanishes except in degrees $0$ and $d-2$.  The latter wont be of much interest to us as we are just interested in the degree zero and one  stretch of the long exact sequence.   For degree 0, it  is nontrivial when $n\geq 0$ where it is given by symmetric trace free tensors with $n$ indices.  Thus
\be\label{les}
0\rightarrow H^0(PT^*_NM,L^n)\stackrel{D_0}\rightarrow  H^0(PT^*_NM , L^{n+1}) \stackrel{\delta}\rightarrow H^1(\PA ,L^n) \rightarrow  0\, .
\ee

The connecting homeomorphism $\delta$ at degree zero to one thus  gives the isomorphisms
\be\label{cohomology}
H^1(\PA ,L^n)= H^0(PT^*_NM , L^{n+1})/D(H^0(PT^*_NM,L^n))
\ee
and this is equivalent to \eqref{Ptrans} by contraction of the tensors on the right hand side of \eqref{Ptrans} with $n+1$ copies of $P$.  Since $P$ is null, we can only determine trace-free symmetric tensors from their contractions with copies of $P$. $\Box$

\smallskip

In particular, for $n=0$, we obtain off-shell Maxwell fields modulo gauge, and for $n=1$ we obtain linearized trace-free metrics (the trace-free condition means that we are really just talking about conformal structures) modulo diffeomorphisms.  

It is instructive to see how the transform works explicitly in terms of the Dolbeault representatives we will use.  We will just work through the $n=0$ case as all the others work very similarly.  Starting from space-time, we will have a Maxwell field $A=A_{\mu} dX^\mu$  on $M$.  We can then attempt to find a $P$ dependent gauge transformation $a(X,P,\bar P)$ so that $A-d \alpha$ descends to $P\A{}$.  Thus we must solve
\be\label{gauge}
P^\mu\frac{\p \alpha}{\p X^\mu}=P^\mu A_\mu\, .
\ee
It is always possible to find a solution $\alpha$ to to this equation holomorphically in $P$ locally.  However, if it were holomorphic in $P$ globally, it would, by Liouville's theorem, be independent of $P$ and would represent a gauge transformation to  the zero Maxwell field.  Thus it must depend nonholomorphically on $P$, but we can nevertheless assume that it will be holomorphic in $X$ as we work on an analytically trivial subset of complex space-time.    We then define 
\be
a:=\frac {\p \alpha}{\p \bar P_\mu} d\bar P_\mu \in H^1_{\delbar} (\PA,\cO)\, .
\ee
The fact that $a$ descends to $P\A{}$ follows by acting on \eqref{gauge} with $\p/\p \bar P_\mu$ and its $\delbar$ closure follows from its $\delbar$ closure (indeed exactness) on $\PT^*M$.  

In the converse direction, given such an $a$, we can pull it back to $PT^*_N M$.  On the fibres, $a$ must be cohomologically trivial and so can be expressed as $a  =d\alpha$ for some $\alpha$.  Since $a$ is pulled back from $P\A{}$, we have  $\cL_{D_0} \alpha=0$ and this yields  $\cL_{D_0}\delbar \alpha=\delbar D\alpha=0$.  Thus  $D_0\alpha$ is holomorphic in $P$ and $X$ globally in $P$ and so by Liouville's theorem in the $P$ variables adapted to homogeneity degree-1, $D_0\hook a=A_\mu P^\mu$ for some $A_\mu$.  

 For the case of a momentum eigenstate, $A=\e^{ik\cdot X} \epsilon_\mu dX^\mu$  we see that the above chain of correspondences is fulfilled by
\be\label{gaug-trans}
\alpha = \frac{\epsilon\cdot P}{k\cdot P} \e^{ik\cdot X}\, , \quad \mbox{ so } \quad a= \e^{ik\cdot x} \epsilon\cdot P \delbar \frac1{k\cdot P}\, .
\ee 
For a complex variable $z$, $\delbar\frac 1z$ is a  distributional $(0,1)$-form $\bar \delta (z)$ with delta function support at $z=0$ so we may write
\be\label{dolrep}
 a= \e^{ik\cdot x} \epsilon\cdot P \, \bar\delta(k\cdot P)\, .
\ee
We remark that on the support of the delta function $D_0 \e^{ik\cdot X}=0$ so it is clear that this representative descends to $\PA$.  This is defined irrespective of whether $k^2$ vanishes or not.

The Penrose transform for $n=0, 1$ has nonlinear extensions.  The case $n=0$ coresponds directly to a deformation of the complex structure on the trivial line bundle and naturally extends to nonabelian Yang-Mills fields: given a bundle $E'$ with connection $A$ on $M$, we can define a holomorphic bundle $E\rightarrow \PA$ whose fibre at a null geodesic $n\in \PA$ is the space of covariantly constant sections of $E'$ over the corresponding null geodesic.  It can be seen that $(E',A)$ can be reconstructed from $E$ as a holomorphic vector bundle and the correspondence is stable under small deformations.  Thus any holomorphic vector bundle on $P\A{}$ that is a deformation of the trivial bundle will give rise to a Yang-Mills field on space-time 

In the case $n=1$, given $h\in H^1(P\A{},L)$, we  can construct the corresponding Hamiltonian vector field $X_h$ with respect to the symplectic structure $\omega$ yielding $X_h\in H^1(\PA,T^{1,0}P\A{})$.  It thus corresponds to the infinitesimal deformation of the complex structure $\delbar \rightarrow \delbar + X_h$.   By construction it preserves the existence of the holomorphic symplectic structure.  It also preserves the existence of the holomorphic contact structrure as 
\be
\cL_{X_h}\theta=d\theta(X_h) +\omega(X_h,\cdot)=dh-dh=0\, .
\ee
since we have from the Euler relation
\be\label{recon-h}
\omega(\Upsilon,\cdot)=\theta(\cdot)\,  , \qquad  \theta(X_h)=\omega(\Upsilon,X_h)=\Upsilon (h)=h\, .
\ee
Thus, this is a linearized deformation of the complex structure of $\PA$ that preserves the holomorphic contact and symplectic structures on $A$, and we see from the above that this corresponds precisely to variations of the conformal structure of $M$, see \cite{Baston:1987av} for the 4-dimensional case.  

We can understand the role of $h$ more directly by observing that the contact structure determines the complex structure.  This is because $d\theta$ is nondegenerate on $T^{1,0}\A$ and so determines  $T^{0,1}\A$ as those complex vector fields that annihilate $d\theta$.  Under the deformation determined by $h$, the deformed contact structure is $\theta_h=\theta-h$ to first order as $\theta_h$ must annihilate the deformed $\delbar$-operator $\delbar_h=\delbar+X_h$ and as we have seen $\theta(X_h)=h$.  Thus $h$ is the deformation of $\theta$.

\subsection{The heterotic extension}\label{P-Transform-RNS}
 
We will now take $\A$ to be the supersymmetric ambitwistor space 
appropriate to the heterotic case of dimension $(18|8)$ (the type II
version is $(18|16)$ dimensional).

We again construct this super-ambitwistor space to be symplectic reduction.
We  extend the cotangent bundle coordinates $(X,P)$  with the $d$
fermionic coordinates $\Psi^\mu$ and the symplectic potential and 2-form by
\be\label{susy-symp}
\theta=P_\mu dx^\mu + g_{\mu\nu}\Psi^\mu d\Psi^\nu/2\, , \qquad \omega = d\theta= dP_\mu\wedge dX^\mu + g_{\mu\nu} d\Psi^\mu d\Psi^\nu/2\, .
\ee
We now perform the symplectic reduction by both $P^2$ and $P\cdot\Psi$.  Thus we set $P^2=P\cdot\Psi=0$ and quotient by $D_0=P\cdot \nabla $ and now also  $D_1=\Psi\cdot \nabla + P\cdot \p/\p \Psi $. 
Thus we can define $\A$ to be the quotient of the bundle $T^*_{SN}M$ of super null vectors as follows
\be
\A=T_{SN}^*M/\cD \, , \quad \mbox{ where }\quad T^*_{SN}M=\left\{ (X,P,\Psi)\in T^*\oplus \Pi TM \, |\, P^2=0=P\cdot \Psi\right\} 
\ee
where $\cD$ is the distribution given by
\be
\cD:=\{D_0,D_1\}:=  \left\{P\cdot \nabla\, ,   \Psi\cdot \nabla + P\cdot \frac \p{\p \Psi}\right\}\, .
\ee

For the projectivisation $\PA$, we take the quotient by the Euler vector 
\be\label{proj}
\Upsilon=2P\cdot \frac\p{\p P} +  \Psi\cdot \frac \p{\p\Psi}\, , \qquad \PA= \A/\Upsilon \, ,  
\ee
so that, before the quotient by $\cD$ we are taking the equivalence relation $(X,P,\Psi)\sim (X, \lambda^2 P,\lambda \Psi)$ making the fibres of $PT^*_{SN}M\rightarrow M$ a weighted projective super space.  It is easy to see that $\Upsilon$ preserves  $\cD$ and descends to $\A$ and so expresses $\A$ as the total space of a line bundle $\cO(-1)\rightarrow \PA$ with $P$ taking values in $\cO(2)$ and $\Psi$ in $\cO(1)$.  

We can follow the same strategy for the Penrose transform as in the
purely bosonic case.  We will just discuss the low lying examples that
are relevant in detail.  

\begin{thm}
We have that $H^1(\PA,\cO(n))$ vanishes for $n<-1$.   For $n\geq -1$
elements 
correspond to a polynomial  in $(P,\Psi)$ of weight $n+1$ whose
coefficients are arbitrary holomorphic functions of $X$, modulo
$D_1$ of an arbitrary polynomial in $(P,\Psi)$ of degree $n$. 
\end{thm}

The proof follows the strategy given before and can be obtained from
the long exact sequence in cohomology that follows from the short exact sequence 
\be\label{SES1}
0\rightarrow \cO(n)_{\PA } \rightarrow \cO(n)^n_{ PT^*_NM } \stackrel{D_1}{\rightarrow} \cO(n+1)_{PT^*_NM}\rightarrow 0\, .
\ee
This is essentially \eqref{SES} but with $D_0$ replaced by $D_1$.
As before we pull $a\in
H^1(\PA,\cO(n))$ back to $PT^*_{SN}M$ and deduce that on this
space $a=\delbar \alpha$ for some $\alpha(X,P,\Psi)$ of weight $n$,
defined up to the addition of polynomials in $(P,\Psi)$ of weight $n$
whose coefficients are arbitrary functions of $X$ alone.   Because
$D_1 a=0$, $\delbar D_1\alpha=0$ so that $D_1\alpha$ is global and
holomorphic, and hence  a polynomial of degree $n+1$ in $(P,\Psi)$
whose coefficients are arbitrary functions of $X$.  The gauge freedom
in $\alpha$ gives the stated gauge freedom in $D_1\alpha$.

The simplest case is the weight zero case and we will start with a
choice of $a\in H^1(\PA,\cO)$. 
Since the vectors in $\cD$ acting on $a$ vanish, we have that $D_0\alpha$ and $D_1\alpha$ are holomorphic in $P$ and $\Psi$ respectively of weight $2$ and $1$.  We can therefore  expand
\be\label{expand}
 D_1\alpha=\Psi^\mu A_\mu\, .
\ee
Since $D_0=D_1^2 $ we will have 
\be
D_0\alpha= P^\mu A_\mu + \Psi^\mu\Psi^\nu F_{\mu\nu}\, .
\ee
Thus we have an off-shell Maxwell field $A$ defined up to gauge.

The $\alpha$ and $a$ associated to a momentum eigenstate
$A=\e^{ik\cdot X}\epsilon_\mu dX^\mu$ are 
\be\label{susy-dolrep}
\alpha =e^{ik\cdot X}\frac{\epsilon \cdot P + \epsilon\cdot \Psi
  k\cdot \Psi}{k\cdot P}\, , \qquad a=  e^{ik\cdot X} (\epsilon\cdot P + \epsilon\cdot \Psi k\cdot \Psi)\bar\delta( P\cdot k)\, .
\ee

The same strategy can be applied to all $\cO(n)$ albeit with
increasing complexity.  For $\cO(-1)$ it is easy to see that one obtains
a scalar field.  For $\cO(2)$ we obtain a rank two tensor (without any symmetry or trace assumption) and a 3-form
\be\label{O2-susy}
H^1(\PA,\cO(2))=\{H_{\mu\nu},C_{\mu\nu\rho}=C_{[\mu\nu\rho]}\}/\{ \nabla_\mu v_\nu + w_{\mu\nu}, \nabla_{[\mu}w_{\nu\rho]} \}\, .
\ee
where $w_{\mu\nu}=w_{[\mu\nu]}$.
The corresponding Dolbeault representative for such a set of fields of the form $\e^{ik\cdot X} H_{\mu\nu}$ etc., with $H$ and $C$ constant is
\be\label{dol-grav-rep}
h= \e^{ik\cdot X}\bar\delta(k\cdot P) \left( P^\mu P^\nu H_{\mu\nu} - P^\mu\Psi^\nu\Psi^\rho (H_{\mu\nu}k_\rho + 3 C_{\mu\nu\rho})-\Psi^\mu\Psi^\nu\Psi^\rho\Psi^\sigma k_\mu C_{\nu\rho\sigma}\right)
\ee
As in the weight zero case, the pullback of $h$ to $PT_{SN}^*M$  is trivial with $h=\delbar \eta$ where
\be
\eta=\frac{\e^{ik\cdot x}}{k\cdot P} \left( P^\mu (P^\nu +k\cdot \Psi\Psi^\nu)H_{\mu\nu}-(3P^\mu+k\cdot\Psi\Psi^\mu)\Psi^\nu\Psi^\rho C_{\mu\nu\rho}\right) 
\ee
and we have
\begin{eqnarray}\nonumber
D_0\eta&=&   \left(P^\mu (P^\nu +k\cdot \Psi\Psi^\nu)H_{\mu\nu}-(3P^\mu+k\cdot\Psi\Psi^\mu)\Psi^\nu\Psi^\rho C_{\mu\nu\rho} \right)\e^{ik\cdot x}\\
D_1\eta&=& \left(P^\mu  \Psi^\nu H_{\mu\nu}+ \Psi^\mu\Psi^\nu\Psi^\rho C_{\mu\nu\rho}\right)\e^{ik\cdot x}\, .
\end{eqnarray}
We can interpret these as determining linearized deformations of the constraints underlying the symplectic reduction, with the first representing a deformation of $P^2$ and the second of $P\cdot \Psi$.  The gauge freedom can be seen to arise from diffeomorphisms of $PT^*_{SN}M$ generated by Hamiltonian vector fields of functions of the form $P\cdot v+\Psi^\mu\Psi^\nu w_{\mu\nu}$ which corresponds to the natural Lie lift of a vector field on $M$ together with an infinitesimal rotation of the $\Psi^\mu$.

As before, unlike Witten's super-ambitwistor construction in 10 dimensions \cite{Witten:1985nt}, our fields $A$, $h$, $C$  are completely off-shell.  The on-shell conditions will arise from quantum corrections  to the BRST invariance that corresponds to the quotient by $\cD$.  These will correspond to the application of second order operators $\nabla\cdot \nabla$ and $\nabla\cdot\p/\p\Psi$ to the representatives above.  It is straightforward to see that, as a combination, these operators descend to $\PA$ and so can be consistently applied to $\alpha$ and $\beta$.  The first of these simply gives $k^2=0$ so that $k$ is null.  The second gives $k\cdot \epsilon=0$ for $\alpha$ and $H_{\mu\nu}k^\nu=0$ together with $k^\mu C_{\mu\nu\rho}=0$ for $\beta$.

Again there are non-linear extensions of these transforms as in the bosonic case.  The case of $H^1(\PA,\cO)$ extends naturally to give an encoding of Yang-Mills fields on space-time in terms of holomorphic vector bundles on $\PA$ that are deformations of the trivial  bundle.  Similarly, $h\in H^1(\PA,\cO(2))$ naturally corresponds to deformations of the contat structure $\theta_h=\theta-h$ that determines the complex structure as in the bosonic case.  The Hamiltonian vector fields using the supersymmetric extension of the symplectic structure of members of $H^1(\PA, \cO(2))$ give a direct representation of the associated complex structure deformation;   these are the deformations of the complex structure of $\PA$ that preserve the symplectic potential and symplectic structure.  
It  would be interesting to understand how the on-shell conditions can be imposed in the non-linear regime.


\end{document}